\documentclass[trackchanges]{aastex701}

\usepackage{amsmath}
\usepackage{makecell}
\usepackage{bm}
\usepackage{array}
\usepackage[table]{xcolor}
\newcommand{\vect}[1]{\bm{#1}}

\begin{document}

\title{JUG: JAX-based Unified pulsar timinG}

\author[gname=Matthew, sname='Miles']{M.~T.~Miles}
\affiliation{Department of Physics \& Astronomy, Vanderbilt University, 2301 Vanderbilt Place, Nashville, TN 37235, USA}
\affiliation{OzGrav: The ARC Center of Excellence for Gravitational Wave Discovery, Hawthorn VIC 3122, Australia}
\email[show]{matthew.t.miles@vanderbilt.edu}  

\author[gname=Stephen, sname='Taylor']{S.~R.~Taylor} 
\affiliation{Department of Physics \& Astronomy, Vanderbilt University, 2301 Vanderbilt Place, Nashville, TN 37235, USA}
\email{stephen.r.taylor@Vanderbilt.Edu}

\author[gname=Matthew, sname='Bailes']{M.~Bailes}
\affiliation{Centre for Astrophysics and Supercomputing, Swinburne University of Technology, Hawthorn VIC 3122, Australia}
\affiliation{OzGrav: The ARC Center of Excellence for Gravitational Wave Discovery, Hawthorn VIC 3122, Australia}
\email{mbailes@swin.edu.au}

\author[gname=Aurelien, sname='SD']{A.~Chalumeau}
\affiliation{LPC2E, OSUC, Univ Orleans, CNRS, F-45071 Orleans, France}
\email{aurelien.chalumeau@cnrs-orleans.fr}

\author[gname=Thankful, sname='Cromartie']{H.~T.~Cromartie}
\affiliation{Department of Physics \& Astronomy, Vanderbilt University, 2301 Vanderbilt Place, Nashville, TN 37235, USA}
\email{thankful.cromartie@vanderbilt.edu}

\author[gname=Kyle, sname='Gersbach']{K.~ A.~ Gersbach}
\affiliation{Department of Physics \& Astronomy, Vanderbilt University, 2301 Vanderbilt Place, Nashville, TN 37235, USA}
\email{kyle.a.gersbach@vanderbilt.edu}

\author[gname=Rutger, sname='van Haasteren']{R.~van~Haasteren}
\affiliation{Max Planck Institute for Gravitational Physics (Albert   Einstein Institute), Leibniz Universit\"at Hannover, Callinstrasse 38,   D-30167, Hannover, Germany}
\email{rutger@vhaasteren.com}

\author[gname=Michael, sname='Keith']{M.~J.~Keith}
\affiliation{Jodrell Bank Centre for Astrophysics, Department of Physics and Astronomy, University of Manchester, Manchester M13 9PL, UK}
\email{michael.keith@manchester.ac.uk}

\author[gname=Nima, sname='Laal']{N.~Laal}
\affiliation{Department of Physics \& Astronomy, Vanderbilt University, 2301 Vanderbilt Place, Nashville, TN 37235, USA}
\email{nima.laal@vanderbilt.edu}

\author[gname=Michael, sname='Lam']{M.~T.~Lam}
\affiliation{SETI Institute, 339 N Bernardo Ave Suite 200, Mountain View, CA 94043, USA}
\email{michael.lam@nanograv.org}

\author[gname=Kuo, sname='Liu']{K.~Liu}
\affiliation{State Key Laboratory of Radio Astronomy and Technology, Shanghai Astronomical Observatory, CAS, 80 Nandan Road, Shanghai 200030, P. R. China}
\email{liukuo@shao.ac.cn}

\author[gname=Aditya, sname='Parthasarathy']{A.~Parthasarathy}
\affiliation{ASTRON, Netherlands Institute for Radio Astronomy, Oude Hoogeveensedĳk 4, 7991 PD Dwingeloo, The Netherlands}
\affiliation{Anton Pannekoek Institute for Astronomy, University of Amsterdam, Science Park 904, 1098 XH Amsterdam, The Netherlands}
\affiliation{Max-Planck-Institut f\"{u}r Radioastronomie, Auf dem H\"{u}gel 69, 53121, Bonn, Germany}
\email{parthas@astron.nl}

\author[gname=Scott, sname='Ransom']{S.~M.~Ransom}
\affiliation{National Radio Astronomy Observatory, 520 Edgemont Rd., Charlottesville, VA 22903}
\email{sransom@nrao.edu}

\author[gname=Daniel, sname='Reardon']{D.~J.~Reardon}
\affiliation{Centre for Astrophysics and Supercomputing, Swinburne University of Technology, Hawthorn VIC 3122, Australia}
\affiliation{OzGrav: The ARC Center of Excellence for Gravitational Wave Discovery, Hawthorn VIC 3122, Australia}
\email{dreardon@swin.edu.au}

\author[gname=Ryan, sname='Shannon']{R.~M.~Shannon}
\affiliation{Centre for Astrophysics and Supercomputing, Swinburne University of Technology, Hawthorn VIC 3122, Australia}
\affiliation{OzGrav: The ARC Center of Excellence for Gravitational Wave Discovery, Hawthorn VIC 3122, Australia}
\email{rshannon@swin.edu.au}

\author[gname=David, sname='Wright']{D.~C.~Wright}
\affiliation{Department of Physics, Oregon State University, Corvallis, Oregon 97331, USA}
\email{wrightd2@oregonstate.edu}

\author[gname=Andrew, sname='Zic']{A.~Zic}
\affiliation{Australia Telescope National Facility, CSIRO, Space and Astronomy, PO Box 76, Epping, NSW 1710, Australia}
\affiliation{OzGrav: The ARC Center of Excellence for Gravitational Wave Discovery, Hawthorn VIC 3122, Australia}
\email{andrew.zic@csiro.au}


\begin{abstract}
We present \textsc{JUG} (JAX-based Unified pulsar timinG), a \textsc{JAX}-based, fully independent pulsar timing package emphasising speed and ease of use, designed to confidently handle the increasingly large and complex pulsar timing array datasets that are being created in the pulsar timing field. \textsc{JUG} implements the entire pulsar timing pipeline itself, from data handling and clock corrections through to the timing model and fitting, without relying on other timing software. It enables Pythonic programming at the speed of compiled code, is GPU-capable, and can be operated via a Python API or an interactive GUI. A user of \textsc{JUG} can interactively explore data, fit timing models with complex stochastic noise, model deterministic signals such as continuous gravitational waves, and obtain accurate point estimates of the parameters of the stochastic processes present, thereby bridging frequentist timing and Bayesian noise analysis. \textsc{JUG} is faster than \textsc{PINT} by more than fifty times and is comparably fast to \textsc{Tempo2}, can handle millions of arrival times, agrees with \textsc{PINT} at the picosecond level, and can reliably recover known timing model and noise parameter values. In this paper we describe its design, performance, and validation, and demonstrate its advantages for pulsar timing data analysis. 
\end{abstract}

\keywords{Pulsars, Millisecond Pulsars, Astronomy Software}


\section{Introduction} 
Since their discovery \citep{1968Natur.217..709H} pulsars have been responsible for a litany of scientific breakthroughs. The study of their emission, particularly the emission of millisecond pulsars \citep{1982Natur.300..615B, 1982Natur.300..728A}, has been used for tests of general relativity \citep{1982ApJ...253..908T, 2003Natur.426..531B, Kramer_double_pulsar_2006}, led to the discovery of planetary systems \citep{Wolszczan_Frail_1992, Wolszczan_planets_1994}, and put constraints on nuclear equations of state \citep{Demorest_Nature_2010, Antoniadis1233232, 2021ApJ...915L..12F, 2021arXiv210506979M, 2021arXiv210506980R, Cromartie_Nature_2020}. More recently, long-term pulsar timing experiments have led to results detailing the first evidence for a gravitational wave background \citep{2023RAA....23g5024X, 2023arXiv230616214A, 2023ApJ...951L...8A, 2023ApJ...951L...6R, 2025MNRAS.536.1489M}, and it is expected that these methods will soon be used to directly detect a discrete source of nanohertz-gravitational waves \citep{2015MNRAS.451.2417R, 2018MNRAS.477..964K, 2022ApJ...941..119B}. 

These discoveries have had significant impact on the astronomy field, but they are also all drawn from a single mechanism of analysis: that of pulsar timing \citep{1989ASIC..262...17T, bailes_2009}. By this method, pulsed emission incoming from distant pulsars is cross-correlated against high-fidelity representations of the pulse and ``timed", their arrival to the Earth predicted in some cases to nanoseconds. Analysis of this emission leads to the discoveries previously mentioned, but few software options exist for this type of work, described entirely by \textsc{Tempo}, \textsc{Tempo2}, and \textsc{PINT} (earlier codes such as \textsc{psrtime}\footnote{\url{https://www.jb.man.ac.uk/~pulsar/observing/progs/psrtime.html}} also saw wide use). These packages have been more than capable for the analysis of pulsar timing data using current datasets, however, pulsar timing array (PTA) datasets are rapidly increasing in volume. With new facilities such as the Deep Synoptic Array \citep{2019BAAS...51g.255H} and the SKA \citep{2009IEEEP..97.1482D} soon to be commissioned, PTA datasets in the gravitational-wave detection era are including more arrival times, more pulsars, and more complex modelling to achieve more precise pulsar timing. The combination of these strains existing tools, necessitating a new architecture that is built to handle these more complicated data. 

\textsc{Tempo} \citep{2015ascl.soft09002N} and \textsc{Tempo2} \citep{2006MNRAS.369..655H} are commonly used, fast, compiled options to handle pulsar timing data. However, they are not easily extensible and are written in Fortran and C/C++ respectively, languages that are less widely used for analysis tasks in the astronomy community than Python. A more recent successor to these is \textsc{PINT} \citep{2021ApJ...911...45L}, an independent Python implementation of the pulsar timing algorithms. This software is very extensible and easy to understand, however, it does not scale well with large numbers of arrival times in pulsar datasets or with complicated models. Finally, there is also \textsc{Vela} \citep{2025ApJ...980..165S}, a recently developed independent timing model software written in the Julia language, focused on enabling Bayesian pulsar timing, relying on \textsc{PINT} only for some data handling tools. To try and tackle these pressure points in the existing software, we introduce \textsc{JUG}, a fully independent timing software written in Python and built prioritising the handling of large datasets at speed using \textsc{JAX} \citep{jax2018github}.

\textsc{JUG} makes use of the \textsc{JAX} software package to just-in-time (JIT) compile and fuse the timing model delays into a single kernel evaluated over all the vectorised arrival times at once. This allows the speed of \textsc{JUG} to reach that of a compiled codebase, but written in Python, resulting in a package that is more than an order of magnitude faster than \textsc{PINT} and comparable to \textsc{Tempo2}'s fast \texttt{QR} solver, while scaling at speed to large datasets ($10^6$ arrival times). \textsc{JUG} has been developed to operate completely independently of other timing software. This independence is of the software itself rather than its development process, and \textsc{PINT} in particular was used as a reference implementation against which \textsc{JUG} is validated (Section~\ref{sec: validation}). Further, \textsc{JUG} does not operate only as a forward model evaluator. It also fits timing parameters through weighted or generalised least-squares solver routines (WLS/GLS) optimised for speed, and is able to quickly estimate parameters of common pulsar noise processes through a fast maximum-a-posteriori point estimate method, bridging the gap between frequentist timing analysis and Bayesian noise parameter estimation. \textsc{JUG} is able to model various additional deterministic signals that are not usually included in the pulsar timing model, and \textsc{JAX} natively exposes auto-differentiation for forward modelled elements that may soon enable alternative fitting mechanisms. Further, \textsc{JUG} is designed to be operated according to user preference, allowing those more familiar with the use of a \textsc{Tempo2} style graphical user interface (GUI) to operate \textsc{JUG} through this manner, or those more familiar with \textsc{PINT}'s programmatic style of data analysis to use the \textsc{JUG} API. In this paper we provide a brief background on the art of pulsar timing (Section \ref{sec: Background on timing}), the design and implementation of the \textsc{JUG} software package (Section \ref{sec: design and imp}), the performance of \textsc{JUG} (Section \ref{sec: performance}), its validation compared to other timing software (Section \ref{sec: validation}), and a summary and planned future steps for its continued development (Section \ref{sec: summary}).

\section{A Background on Pulsar Timing} 
\label{sec: Background on timing}
In traditional pulsar timing data analysis, collecting and interpreting the time of arrival (TOA) of the pulse from the pulsar is the principal aim. Formally, this TOA is defined as the observable time that the pulse from the pulsar arrives at the observing station on the Earth. In practice, pulsar timing relies on averaging many individual pulses together and assigning that average a corresponding arrival time. This is necessary because pulsar timing assumes the pulse is a stable, well-averaged representation of the pulse-energy distribution through phase. After averaging many pulses together, the integrated profile converges to a stable shape as a consequence of the central limit theorem \citep{1975ApJ...198..661H}, while often the individual pulses will appear markedly different from one another. In addition, averaging individual pulses together in the radiometer-noise limit increases the signal-to-noise ratio (S/N) as $\mathrm{S/N} \propto \sqrt{N_p}$, where here $N_p$ represents the total number of pulses averaged together. The most common way to calculate this TOA is by cross-correlating the prepared data products with a high-fidelity representation of the pulsar's average pulse profile \citep{1992RSPTA.341..117T}. This representation is commonly referred to as a template in the case that it is a single-frequency representation, and a portrait where it is resolved through radio frequency \citep{2016ascl.soft06013P}. As mentioned earlier, we assume that the pulse after a degree of averaging can be considered stable through time \citep{2004hpa..book.....L}, and as such any averaged pulse from a particular observation ($\mathcal{O}(t)$) can be measured simply through \citep{bailes_2009}
\begin{equation}
\label{eq: phase_shift_eq}
    \mathcal{O}(t) = \alpha + \beta\mathcal{P}(t-\tau) + \epsilon(t)
\end{equation}
where $\alpha$ is an arbitrary offset, $\beta$ is a scale factor, $\mathcal{P}(t)$ is the template or portrait, $\epsilon(t)$ is any noise perturbing the profile, and $\tau$ is the shift in time between $\mathcal{O}(t)$ and $\mathcal{P}(t)$. The calculated shift, $\tau$, is estimated through matched-filtering the observation against the template, usually in the Fourier domain. This is ultimately applied to the time of the observation recorded at the telescope site, resulting in a TOA measurement and a corresponding uncertainty derived from the matched-filtering algorithm.

The series of TOAs that can be measured in this manner for a given pulsar must be interpreted from an understanding of the pulsar's underlying physics and the relative environment along the Earth-pulsar line of sight. Facilitating this, a pulsar timing model, or ephemeris, is constructed to describe both the pulsar's rotational phase and the chain of delays that alter the pulse arrival time. Only when these mechanisms are accurately determined to the sensitivity of the data can a pulsar be timed accurately and scientific inquiry be informed. Here we describe the pulsar timing model in brief as it is well established in pulsar timing literature \citep{2006MNRAS.369..655H, 2021ApJ...911...45L} and \textsc{JUG} has followed this, not attempting to improve upon the well established convention.

The rotation of a pulsar is represented in the timing model through the pulsar's rotational phase $\phi(t)$, a frame of reference that is comoving with the pulsar itself expressed as a function of time, $t$. While pulsars are very rotationally stable, their rotational frequency ($f$) will change slowly through time as energy is lost via the emission of the pulse. The degree to which this occurs is unacceptable for the precision desired in pulsar timing experiments, and this is accounted for by expanding the phase as a Taylor expansion in time of the form
\begin{equation}
\label{eq: rot_phase}
    \phi(t) = \phi_0 + f(t-t_0) + \frac{1}{2}\dot{f}(t-t_0)^2 + \frac{1}{6}\ddot{f}(t-t_0)^3 + ...
\end{equation}
where $\phi_0$ is the initial phase of the pulsar at time $t_0$, and the dots above $f$ denote the degree of time derivative applied to the rotational frequency.

\autoref{eq: rot_phase} refers to the rotational phase as a function of time in a frame comoving with the pulsar (from here $t_\mathrm{psr}$), however, the arrival time measured as a consequence of \autoref{eq: phase_shift_eq} is measured on Earth against a clock at the observatory (from here $t_\mathrm{topo}$). These are not equivalent, and in order to evaluate the pulsar phase, $t_\mathrm{topo}$ must be transformed to the same comoving frame as $t_\mathrm{psr}$ through the aforementioned timing model delay chain $\Delta$
\begin{equation}
    t_\mathrm{psr} = t_\mathrm{topo} - \Delta.
\end{equation}
The corrections in the delay chain fall into two distinct classes: a timescale transformation, converting the observatory clock reading onto a uniform coordinate time, and a transformation of the reference frame from the moving topocenter to the Solar System Barycenter (SSB), an approximation to an inertial frame in which the phase model of \autoref{eq: rot_phase} applies. The total delay $\Delta$ comprises contributions from the clock chain, the SSB transformation, the Ionised Interstellar Medium (IISM), and, for binary systems, the pulsar's orbital motion,
\begin{equation}
    \Delta = \Delta t_\mathrm{clock} + \Delta t_\mathrm{SSB} + \Delta t_\mathrm{IISM} + \Delta t_\mathrm{B},
\end{equation}
which must be applied in this order, as each correction is evaluated given the correction in its preceding term. These effects constitute the dominant contributions to $\Delta$ and are intended to be illustrative rather than exhaustive. The complete set of deterministic modeling components that are implemented in \textsc{JUG} is given in Appendix \ref{app: parameters}.

\subsection{Timescale Transformation}

The timescale adjustment $t_\mathrm{clock}$ must be performed as each pulse is measured relative to a local clock at the observatory. While highly stable, the clocks used by observatories (commonly Hydrogen masers) drift over long periods of time, accumulating offsets of tens to hundreds of nanoseconds over months. Due to this, the TOA must instead be defined at a globally agreed upon time standard, known as Terrestrial Time (TT). The clocks maintained at observatories are monitored via the Global Positioning System (GPS), providing a transfer to a common standard expressed in UTC. This timescale is not continuous, as it contains discrete offsets (leap seconds) to align with the rotation of the Earth. This is adjusted for by converting the timescale into International Atomic Time (TAI), which is a continuous timescale formed by the average of many atomic clocks globally. Finally, this is transferred into TT with a fixed offset. In practice, this is done by using a retrospective realisation, correcting for imperfections in TAI and providing the most uniform standard possible.\footnote{During this process, a small additional delay is accounted for from the propagation of the signal through the atmosphere above the observatory that is dependent on the elevation of the observatory. We denote these as tropospheric effects, and while \textsc{JUG} does model them, we will not go into detail here.} This timescale is still defined at Earth, however, and it must still be transformed to the pseudo-inertial reference frame of the SSB.

\subsection{Solar System Barycenter Transformation}
The corrections to the SSB are the largest that are performed in the delay chain, reaching hundreds of seconds and varying annually with the Earth's orbital motion. These delays can generally be separated into three portions: the Roemer delay, the solar system Shapiro delay, and the solar system Einstein delay \citep{1986AIHPA..44..263D, 1986ARA&A..24..537B}.

The Roemer delay accounts for the light-travel time between the Earth and the SSB projected onto the line of sight to the pulsar. This is heavily dependent on an accurate solar system ephemeris (e.g. the DE series from JPL \citep{2021AJ....161..105P}), and the pulsar's astrometry. The solar system Shapiro delay is the general-relativistic delay accumulated by the pulse's passage through the gravitational potential of the Sun and the planets. The solar system Einstein delay accounts for the time dilation and gravitational redshift relating TT to a barycentric coordinate time. Two coordinate times are commonly used in pulsar timing: Barycentric Coordinate Time (TCB), and Barycentric Dynamical Time (TDB) \citep{2006MNRAS.369..655H}. TCB is defined as a time running at an unscaled Barycentric rate, whereas TDB is rescaled to be close to TT. The difference between the two standards amounts to a constant rate factor, which results in parameters measured in either system differing by corresponding scale factors. \textsc{JUG} natively operates in TDB, but ingests and converts TCB-defined models internally. 

\subsection{Propagation and Chromatic Delays}

The pulsed emission interacts with free electrons as it travels through the IISM, altering the phase and propagating speed of the radio waves; this causes delays to the TOAs as a function of the radio-frequency of the pulse. This phenomenon is known as dispersion, and is corrected to a degree by measuring the dispersion measure (DM), the column density of electrons along the Earth-pulsar line of sight, and adjusting for its expected effect on the pulse. This can be done as the effect is known a-priori, with the delay of the pulse relative to its frequency expected to be
\begin{equation}
\label{eq: DM delay}
    \delta t_\mathrm{DM} = \left(\frac{\mathrm{DM}}{\kappa}\right)\nu^{-2}
\end{equation}
where $\kappa$ is a dispersion measure constant. There are conflicting dispersion measure constants in use in the pulsar timing field, \textsc{JUG} makes use of $\kappa = 2.41 \times10^{-4}\mathrm{MHz}^{-2}\mathrm{cm}^{-3}\mathrm{pc}\ \mathrm{s}^{-1}$ \citep{2004hpa..book.....L}. Accurately constraining this is of paramount importance to accurate timing, and heavily motivates the use of multi-frequency data as standard practice in pulsar timing.

Through time, it is not expected that the DM should remain constant and instead it is expected to change as the line of sight sweeps through the turbulent, inhomogeneous IISM. These variations are often modeled as a smooth deterministic component, captured by a Taylor expansion similar to that of the rotational phase expressed in \autoref{eq: rot_phase}, together with a stochastic component typically expressed as a Gaussian process. Alternatively, the use of a piecewise fit to DM (e.g. DMX) is common in pulsar timing applications, capturing the total variation. However, using this method can lead to difficulties in modeling other radio frequency dependent (otherwise known as chromatic) features in pulse arrival times. 
A further dispersive contribution arises from the solar wind \citep{2007MNRAS.378..493Y}. As the Earth-pulsar line of sight passes near the Sun, the ionised solar plasma results in chromatic delays that are taken into account in the pulsar timing model. These features have the same chromaticity as signatures from DM, and are the largest where the solar angle (the angle between the Earth-pulsar line of sight and the Sun) is at a minimum, allowing them to be separable to a degree. A distinct chromatic effect arises from multipath propagation through the IISM, which produces a steep frequency dependent ($\sim\nu^{-4}$) pulse broadening effect \citep{2004ApJ...605..759B}, inducing a chromatic delay that is distinct from dispersion. \textsc{JUG} does not fit for these as timing delays, and instead assumes that frequency resolved timing methods and further modeling of scattering noise processes will account for this. 

Finally, not all chromatic structure originates in propagation. Frequency-dependent (FD) parameters account for the evolution of the pulse shape through radio frequency. This exists in \textsc{JUG} as a phenomenological polynomial that is fit for in $\log\nu$ \citep{2015ApJ...813...65N}. Although these are highly covariant with dispersive delays, FD parameters are not accounting for propagation delays but instead intrinsic shape changes of the pulsar's pulse.

\subsection{Binary Delays}

The preceding transformations and delays allowed for the pulsar to be interpretable at the SSB, however, a pulsar in a binary possesses a dynamic emission point. In order to appropriately reference the emission to a comoving frame we must account for that binary motion. The major impact of the binary motion can be accounted for in much the same way as the transformation to the SSB, through three main terms: the orbital Roemer delay, the orbital Shapiro delay, and the orbital Einstein delay \citep{1986AIHPA..44..263D, 2006MNRAS.372.1549E}.

The orbital Roemer delay is once again the term with the largest impact. Modeling it accounts for the light travel time across the pulsar's orbit as projected onto the Earth-pulsar line of sight, parameterised by Keplerian orbital parameters. The orbital Shapiro delay accounts for the additional light-travel time of the pulse as it moves through the gravitational potential of the companion. This term, where it is measurable, is of great interest to pulsar astronomers as it can be used to determine the orbital inclination and companion mass \citep{1986AIHPA..44..263D, Demorest_Nature_2010}. The orbital Einstein delay accounts for the time-dilation and gravitational redshift imparted through the varying speed of the pulsar and impacts from its companion. We note here that this is a very general description of pulsar orbital dynamics, as different geometric configurations of the binary system as observed by Earth will alter what is modeled in the binary. There are several different binary models in use by the pulsar timing community that are better suited to particular orbital configurations. These are commonly, but not exhaustively, determined through the eccentricity of the orbit, and its inclination with respect to the Earth. More details of the models included in \textsc{JUG} can be found in Section \ref{subsec: TM components}. 

\subsection{Timing Residuals}
\label{subsec: timing residuals}
The terms detailed here broadly outline the most impactful transformations and delays informing $\Delta$. Accounting for these, the arrival time can be expressed in terms of $t_\mathrm{psr}$, allowing for the evaluation of \autoref{eq: rot_phase}. For any given TOA, the predicted phase is obtained by evaluating $\phi$ at the corresponding pulsar comoving frame time, $\phi_i\equiv \phi(t_{\mathrm{psr},i})$. Because the pulsar signal is periodically repeating, the number of rotations between any two TOAs is not known a-priori, rather the phase is only known for a single rotation. The construction of a timing model that correctly accounts for the pulsar rotation and derivatives will predict each arrival time within a small fraction of a pulse period, allowing an integer rotation count to be employed unambiguously. The timing residual, the difference between the observed and model-predicted arrival times, is then given by the deviation of the predicted phase from the nearest integer number of rotations, expressed in units of time
\begin{equation}
    \label{eq: timing residual}
    r_i = \frac{\phi_i -\left[\phi_i\right]}{f}
\end{equation}
where $[\phi_i]$ is the integer nearest to $\phi_i$, and $f$ here is the spin frequency of the pulsar, as described in \citet{2021ApJ...911...45L}. From this evaluation, it follows that in cases where there are few observations of a pulsar, or a particularly large gap in said observations, resolving a phase-connected solution can be difficult. 

\subsection{The Timing Likelihood}
\label{subsec: timing likelihood}

The residual defined by \autoref{eq: timing residual} is for a single TOA. Here we define a vector $\vect{r}$, collecting the residuals from all of the TOAs in a dataset ($N_\mathrm{TOA}$), providing a foundation that can be used to estimate further timing model parameters. Many of the timing model parameters are non-linear (e.g. those in the binary model), and so any estimation must be performed by linearising about a reference solution or initial estimate, $\vect{p_0}$. A small perturbation $\vect{\varepsilon} = \vect{p} - \vect{p_0}$ alters the timing residuals by
\begin{equation}
    \label{eq: design matrix}
    \vect{r}(\vect{p}) \approx \vect{r}(\vect{p_0}) - \vect{M}\vect{\varepsilon},
\end{equation}
where the design matrix $\vect{M}$ is constructed of columns defining the partial derivatives of the residuals with respect to each parameter that is being varied. As demonstrated earlier (Section \ref{subsec: timing residuals}), a phase-connected solution will yield residuals that are small fractions of a rotation given the change in the parameter value. It follows then that if the steps from the initial parameter value $\vect{p_0}$ are small, the linear approximation of \autoref{eq: design matrix} is reasonable and any resultant parameter corrections are accurate \citep{2013MNRAS.428.1147V, 2021arXiv210513270T}. This can naturally be optimised by iterating the linearised fit through solving \autoref{eq: design matrix} for $\vect{\varepsilon}$, updating the reference solution to $\vect{p_0}$ + $\vect{\varepsilon}$, which becomes the expansion point for the following iterations until the solution converges under a pre-decided criterion.

How we have treated both the formation of the residuals and the construction of the data model so far has been under the assumption that they are completely deterministic, this is not the case. In practice we expect to observe stochastic structures within the residuals that are correlated in time, or within observing epochs; non-exhaustively, these include spin noise, DM noise, and the stochastic changes in phase and morphology of individual pulses that compound to form jitter noise,  modelled as an epoch-correlated process (ECORR). These also must be accounted for if we are to attempt to measure the timing parameters without bias. Following the standard in the literature \citep{2014MNRAS.437.3004L, 2014PhRvD..90j4012V}, these processes are represented as Gaussian processes in a Fourier basis ($\vect{F}$), the coefficients ($\vect{a}$) of which are assigned zero-mean Gaussian priors with a covariance $\vect{\Phi}$, the diagonal of which encodes the power spectral density (PSD) of each Gaussian process and is governed by a set of hyperparameters (e.g. the amplitude and spectral index of a power-law red noise spectrum). There is also uncorrelated white noise ($\vect{n}$) that we include in the model, accounting for measurement or systematic uncertainties on the data. Following this, the data model of \autoref{eq: design matrix} is represented as
\begin{equation}
    \label{eq: data model}
    \vect{r} = \vect{M}\vect{\varepsilon} + \vect{F}\vect{a} + \vect{n}.
\end{equation}

The hyperparameters that govern these processes are not estimated as part of the timing model fit in \textsc{JUG}, consistent with other timing model software such as \textsc{Tempo2} and \textsc{PINT}, with the exception of \textsc{Vela} which samples the timing model parameters and hyperparameters jointly \citep{2025ApJ...980..165S}. In \textsc{JUG}, these are supplied as a fixed noise model which conditions the estimate of the timing corrections $\vect{\varepsilon}$ and the coefficients of the Fourier sine and cosine basis $\vect{a}$. However, \textsc{JUG} does possess the architecture for fast estimation of these hyperparameters, which we describe in Section \ref{subsec: noise estimation}.

The white noise $\vect{n}$ has a diagonal covariance matrix we define as $\vect{N}$, which is built of the rescaled measured arrival time uncertainties via two parameters generally applied per backend system. These are a multiplicative factor (EFAC), and another term added in quadrature (EQUAD), accounting for additional uncorrelated noise in the data. To keep consistent with the operations in \textsc{PINT}, \textsc{JUG} adopts the convention\footnote{We note that there is an alternate convention to account for the white noise in pulsar timing of the form: $N_{ii} = \mathrm{EFAC}^2\sigma_i^2 + \mathrm{EQUAD}^2$.} that the diagonal entries of $\vect{N}$ are
\begin{equation}
    \label{eq: white noise}
    N_{ii} = \mathrm{EFAC}^2\left(\sigma_i^2 + \mathrm{EQUAD}^2\right),
\end{equation}
where $\sigma_i$ is the unscaled measured uncertainty of the $i$-th TOA from the Fourier domain arrival time measurement. ECORR is also a white-noise process as it is uncorrelated between observing epochs, however it is fully correlated within them. Rather than including it in $\vect{N}$, \textsc{JUG} models ECORR through the reduced-rank formalism used for the time-correlated processes. This is done through the use of an epoch-quantisation matrix, $\vect{U}$, that has columns mapping each observing epoch to its arrival times. This is appended to the basis with the Fourier modes with the per-epoch ECORR variances contained in $\vect{J}$. The equivalent kernel-based model of ECORR, in which the epoch covariance blocks are whitened directly, is also implemented in \textsc{JUG}, however this is not the default. The total covariance of the residuals is
\begin{equation}
    \vect{C} = \vect{N} + \vect{U}\vect{J}\vect{U}^{\mathrm{T}}
             + \vect{F}\vect{\Phi}\vect{F}^{\mathrm{T}}.
\end{equation}

Here, the diagonal $\vect{N}$ captures the uncorrelated white noise, while the time and epoch-correlated processes are gathered into $\vect{F}\vect{\Phi}\vect{F}^{\mathrm{T}}$. Under this covariance, the optimal timing solution is the generalised least-squares (GLS) estimate coinciding with the maximum-likelihood solution of the corresponding Gaussian noise model. However, forming $\vect{C}$ explicitly is an $N_\mathrm{TOA}\times N_\mathrm{TOA}$ operation that becomes prohibitively expensive for large datasets. To navigate this as datasets grow, \textsc{JUG} obtains an equivalent solution by instead estimating the timing corrections $\vect{\varepsilon}$ jointly with the stochastic coefficients $\vect{a}$, minimising the penalised least-squares objective
\begin{equation}
    \label{eq: penalised objective}
    \operatorname*{min}_{\vect{\varepsilon},\,\vect{a}} \; \left\lVert
    \vect{N}^{-1/2}\left(\vect{r} - \vect{M}\vect{\varepsilon} - 
    \vect{F}\vect{a}\right)\right\rVert^2 +
    \vect{a}^{\mathrm{T}}\vect{\Phi}^{-1}\vect{a},
\end{equation}
where $\vect{N}^{-1/2}$ weights each residual by its scaled uncertainty (\autoref{eq: white noise}) and the second term is the Gaussian prior on $\vect{a}$ acting as a penalty on the stochastic coefficients. The minimiser of this objective is equivalent to the marginalised generalised least-squares solution, but is solved as a single augmented linear problem rather than inverting the expensive $\vect{C}$. 

When no correlated noise is present, the stochastic term is removed, $\vect{C}$ is exactly equal to $\vect{N}$ which is diagonal, and \autoref{eq: penalised objective} reduces simply to a weighted least-squares (WLS) fit. At each iteration, the quality of the updated solution is assessed with the marginalised quadratic form $\vect{r}^{\mathsf{T}}\vect{C}^{-1}\vect{r}$, evaluated using a Woodbury identity so that $\vect{C}$ is never formed. The parameter update itself, however, is produced directly by the augmented least-squares solve (Section \ref{subsec: fitting}).

\section{Design and Implementation} 
\label{sec: design and imp}
\subsection{Architecture Overview} 
\label{subsec: architecture}
The core module of \textsc{JUG} is the \texttt{TimingSession} object that defines a pulsar dataset. It contains the input pulsar ephemeris and timing files, the timing model itself, any additional noise model included, and intermediate products needed to manipulate the data that are cached after the first load. The design of this module was built with a single primary goal: to minimise the computational cost of pulsar timing data analysis by caching the operations that remain static across the analysis of a pulsar. These include operations such as: parsing the par and tim files, loading clock corrections, converting to TDB, and evaluating the barycentric geometry. Performing these operations a single time and retaining them allows for all subsequent computations to reuse them rather than rebuilding them each time. 

In practice, on a $35,000$-TOA dataset (PSR~J1909$-$3744 from the NANOGrav 15-yr data release \citep{2023ApJ...951L...9A}), constructing a \texttt{TimingSession} and computing the first set of residuals takes $\sim1.3\,\mathrm{s}$ once \textsc{JAX}'s compilation cache is populated, $\sim0.8\,\mathrm{s}$ of which is a one-time JIT compilation occurring only when the cache is empty. Further residual evaluations use the cached intermediate products, costing only $\sim0.2\,\mathrm{s}$ when forcing recomputation and effectively instantaneous retrieval when the result is unchanged.\footnote{Measured on an Intel Core Ultra 9 285K CPU. See Section \ref{sec: performance} for full benchmarking.}

The caching structure we employ is separated into three tiers. Within a \texttt{TimingSession}, input files, clock corrections, TDB conversion, and parameter-independent barycentric geometries are held in memory and reused across calls. The memory cache is lost when a \texttt{TimingSession} ends, however, to avoid repeating the most expensive computation when the same dataset is reloaded, the barycentric geometry is written to disk. This persistent cache is keyed by a hash of the TDB epochs, observatory coordinates, ephemeris, and the \textsc{Astropy} and \textsc{ERFA} versions, so a stored result is reloaded only for an identical dataset computed in an identical environment where it would be reproduced bit-for-bit. Any change therefore invalidates the cached entry and forces recomputation. If the cache becomes uninterpretable for any reason, the path falls through to the initial computation. Finally, \textsc{JUG} configures \textsc{JAX}'s persistent compilation cache, so that the JIT compilation of its numerical kernels (Section \ref{subsec: TM components}) is retained across process launches and not repeated needlessly.

\begin{figure}
    \centering
    \includegraphics[width=\linewidth]{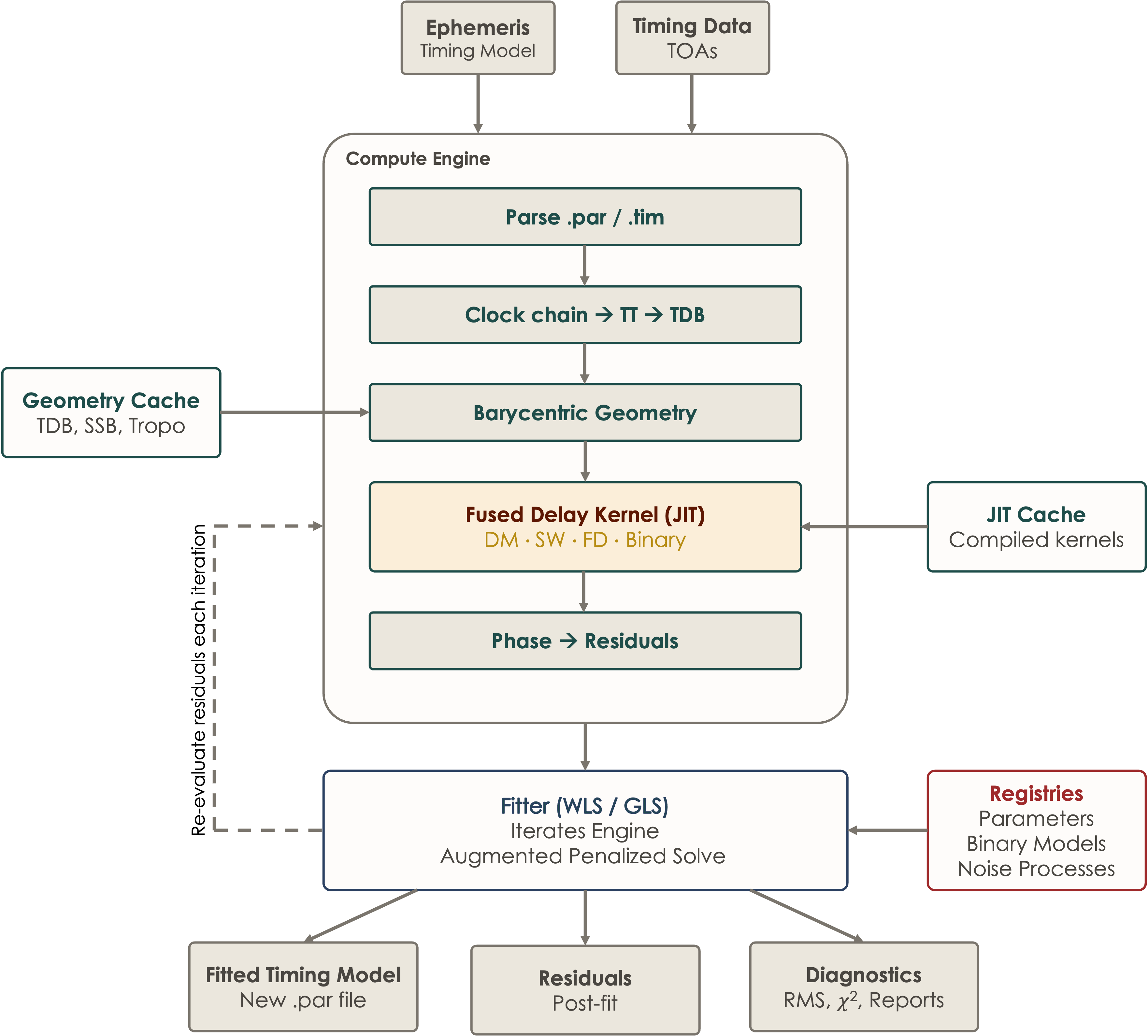}
    \caption{Schematic of the \textsc{JUG} architecture. A pulsar ephemeris (\texttt{.par}) and TOA file (\texttt{.tim}) are taken in by the compute engine, which evaluates the timing model operations: parsing, the observatory clock chain and conversion to TDB, the barycentric geometry, a single fused JIT delay kernel evaluating the dispersion, solar-wind, frequency-dependent, and binary delays, and finally the rotational phase to timing residuals. The geometry and JIT compilations are cached for repeated use. The fitter wraps the engine, re-evaluating the residuals at each iteration of the solver, with the parameter, binary-model, and noise registries checked during fit setup to identify what is present in fitting the model. These output into a new fitted timing model, updated post-fit residuals, and diagnostic summaries for the user.}
    \label{fig: JUG architecture schematic}
\end{figure}

Internally, \textsc{JUG} organises the timing model parameters and the noise processes into registries, with an additional registry holding the set of alternate binary models that can be applied to a pulsar. Each major component of the pulsar model is held in these registries against the parameters they consume and the operations they provide. This allows any relevant set to be assembled automatically from the ephemeris file input. This configuration also means that it can be extended simply without modifying any of \textsc{JUG}'s core architecture. The command-line tools, Python API, and graphical interface (Section \ref{subsec: interfaces}) are all built on the same compute engine. A schematic of this architecture is shown in \autoref{fig: JUG architecture schematic}. 

\subsection{Numerical Precision Strategy} 
\label{subsec: precision}
Accurate pulsar timing necessitates extreme numerical precision on parameters, residuals, and the interpretation of the arrival times in order to infer astrophysical quantities. There are many quantities that need to be held in high precision for accurate timing, however, arguably the most important is that of the rotational frequency and corresponding derivatives. As the rotational phase accumulates as $\phi \approx f(t-t_0)$, over multi-year baselines this can reach extreme values. For a pulsar such as PSR~J1909$-$3744 ($f\approx339\,\mathrm{Hz}$) observed over $\sim15.5$ years in the NANOGrav 15-year dataset, $\phi$ accumulates greater than $10^{11}$ rotations. As the timing residual is the fractional portion of this number, accurately recovering it requires a representation of the phase that is far more precise than simply the integer number of rotations. 

As a software that heavily makes use of \textsc{JAX}, this would raise an issue for \textsc{JUG} were it to naively employ \textsc{JAX} across the architecture as \textsc{JAX} supports precision only up to \texttt{float64} ($15-16$ significant digits). With rotations on the scale of $10^{11}$ as above, this leaves too few remaining significant figures to accurately represent the fractional turn (the timing residual) of the pulsar. The calculated uncertainty on $f$ likewise amplifies across the timing baseline, resulting in a timing error on the order of tens of nanoseconds. This may seem a small number for those unfamiliar with pulsar timing, but it is in fact comparable to the residual precision for some of the most precisely timed pulsars. This then necessitates that a higher precision is held across this parameter in \textsc{JUG}, as well as other similarly compounding parameters. \textsc{JUG} addresses this by using a hybrid precision strategy. It is designed to confine extended precision to the quantities that require it while performing most of the calculations in \texttt{float64}. The rotational frequency and derivatives, the reference epochs (\texttt{PEPOCH}, the binary epochs \texttt{T0} and \texttt{TASC}, and the DM and astrometric epochs) as well as the phase accumulation already mentioned are carried in extended precision (\texttt{longdouble}). These parameters are read directly as strings from the ephemeris into extended precision. The phase is accumulated via Horner's method entirely with \texttt{longdouble} precision, and only the fractional turn is represented in \texttt{float64} before conversion to residuals in seconds. Measured against an arbitrary precision reference calculation using \texttt{mpmath}, \textsc{JUG} reproduces its residuals to an absolute accuracy of $\leq20$ ps over the baseline of PSR~J1909$-$3744. This is the error of \textsc{JUG}'s phase representation against exact arithmetic, and is distinct from any agreement assessed between software packages in Section \ref{subsec: residual agreement}.

In contrast, the chain of propagation and geometric delays is evaluated completely using \texttt{float64} inside the JIT-compiled kernel (Section \ref{subsec: TM components}). This is reasonable as the magnitudes that are used in the calculations are simply the delays referenced to the timing residuals rather than defined at the times of the absolute epochs themselves. The Roemer delay is on the scale of hundreds of ms, the solar-system Shapiro delay is on the order of $\sim100\,\mu s$, and the dispersive delays are on the order of $1$ ms. These magnitudes are held accurately in \texttt{float64}, accumulating a precision uncertainty on the order of $0.1\,\mathrm{fs}$, far lower than attainable timing precision in the foreseeable future. An exception to these is the binary phase which requires a higher precision to accurately constrain the environment of the pulsar binary. In this case, \textsc{JUG} reduces this delay per the orbital period that is kept in extended precision prior to the call to the JIT kernel. The result of this is that only a small fractional in-orbit phase is passed into the lower precision handle. Similarly, the delays are evaluated over lag times formed as $(t_\mathrm{MJD} - t_{0,\mathrm{MJD}}) \times 86400$, differencing in days before scaling to seconds, countering an $\mathcal{O}(100\,\mathrm{ps})$ bias that would otherwise arise.

The division we describe here is completely sufficient for maintaining the required precision and we note that the strategy is well suited to potential hardware acceleration. GPUs do not support extended precision, and should it become beneficial to evaluate residuals on such a unit, the hybrid architecture we describe here is a viable method to do so.

\subsection{Timing Model Components} 
\label{subsec: TM components}

Section \ref{sec: Background on timing} described a chain of delays that are necessary to compute the pulsar's rotational phase from the arrival time collected at an observatory. These delays are usually evaluated in pulsar timing software packages by separate routines and are combined iteratively for each component. This method embraces modularity, but incurs a repeated overhead of dispatching and allocating arrays for each component of the model being solved. \textsc{JUG} sacrifices in part this modularity, evaluating the chromatic and orbital delays within a single fused, JIT compiled kernel \texttt{compute\_total\_delay\_jax}. This fuses these delays into a single compiled function, evaluated over all of the arrival times simultaneously, and reused in subsequent function calls; principally this is the main source of \textsc{JUG}'s per-TOA speed (Section \ref{sec: performance}). A deliberate exclusion from this chain is the optional evaluation of a piecewise dispersion model (DMX), which is instead handled as part of the design matrix as a windowed linear offset. 

In handling binary pulsars, \textsc{JUG} supports the most widely used binary models: \texttt{ELL1} \citep{2001MNRAS.326..274L}, \texttt{DD} \citep{1986AIHPA..44..263D}, \texttt{BT} \citep{1976ApJ...205..580B}, \texttt{DDK} \citep{1995ApJ...439L...5K, 1996ApJ...467L..93K}, and \texttt{T2} \citep{2006MNRAS.369..655H}, as well as their common variants. In these, the orbital Roemer, Shapiro, and Einstein delays are computed together with any post-Keplerian parameters that are included in the model. The \texttt{DDGR} model derives its post-Keplerian parameters from the binary masses under general relativity (GR), and \texttt{DDK} includes the Kopeikin corrections describing contributions of the annual-orbital parallax and the proper motion to the orbital geometry. In an effort to support ephemerides formed from both \textsc{PINT} and \textsc{Tempo2} the general-use \texttt{T2} model is supported but directly dispatched to the appropriate binary model as inferred by the ephemeris parameters. Beyond these components, \textsc{JUG} also models rotational glitches \citep{2026MNRAS.545f2034B}, chromatic exponential events, tropospheric delays, planetary contributions to the solar-system Shapiro delay, and phase jumps that absorb instrumental and system-dependent offsets. Additionally, for convenience, \textsc{JUG} will convert an ephemeris defined in TCB units to TDB internally, evaluating the parameters under the TDB convention. The complete set of supported deterministic components is listed in Appendix \ref{app: parameters}. Further than the timing model capabilities as they stand, \textsc{JUG} also models deterministic signals of astrophysical interest, such as continuous gravitational waves, bursts-with-memory, and chromatic transient events. Given that the waveform models are differentiable within the \textsc{JAX} framework, the architecture supports gradient-based parameter estimation which we plan to develop into full parameter fitting.

\subsection{Fitting} 
\label{subsec: fitting}
In Section \ref{sec: Background on timing} it was established that the fit to the timing model is performed as the minimisation of a penalised least-squares objective (\autoref{eq: penalised objective}). Here, we describe how \textsc{JUG} meets that objective. 

The design matrix $\vect{M}$ per \autoref{eq: design matrix} is assembled from the partial derivatives of the residuals with respect to the parameters being fit at each iteration. These derivatives are exact. \textsc{JUG} evaluates them as analytic expressions for the spin, astrometric, dispersion, frequency-dependent, solar-wind, and jump parameters, as well as the \texttt{DD} and \texttt{ELL1}\footnote{We note that \textsc{JUG} includes the $\cos{\Phi}$ factor in the derivative of the \texttt{ELL1} Shapiro delay with respect to orbital phase, as per \citet{2001MNRAS.326..274L}. This term is absent from the \texttt{ELL1} model of \textsc{PINT}~v1.1.4, although it is present in the \texttt{DD} and \texttt{ELL1H} models. The impact of including this term is almost negligible.} family of binary models. The \texttt{BT} family of binary models are instead evaluated through exact automatic differentiation provided automatically by \textsc{JAX}\footnote{This choice was made as this was a late addition to the binary model tree, it provides an identical solution.}. The analytic derivatives implemented in \textsc{JUG} follow standard conventions established in existing literature as implemented by \textsc{Tempo2} and \textsc{PINT} \citep{2006MNRAS.369..655H, 2021ApJ...911...45L}. Consistent with the other operations in \textsc{JUG}, these are written using \textsc{JAX}, JIT compiled, and vectorised over the set of arrival times.

\begin{figure}
    \centering
    \includegraphics[width=\linewidth]{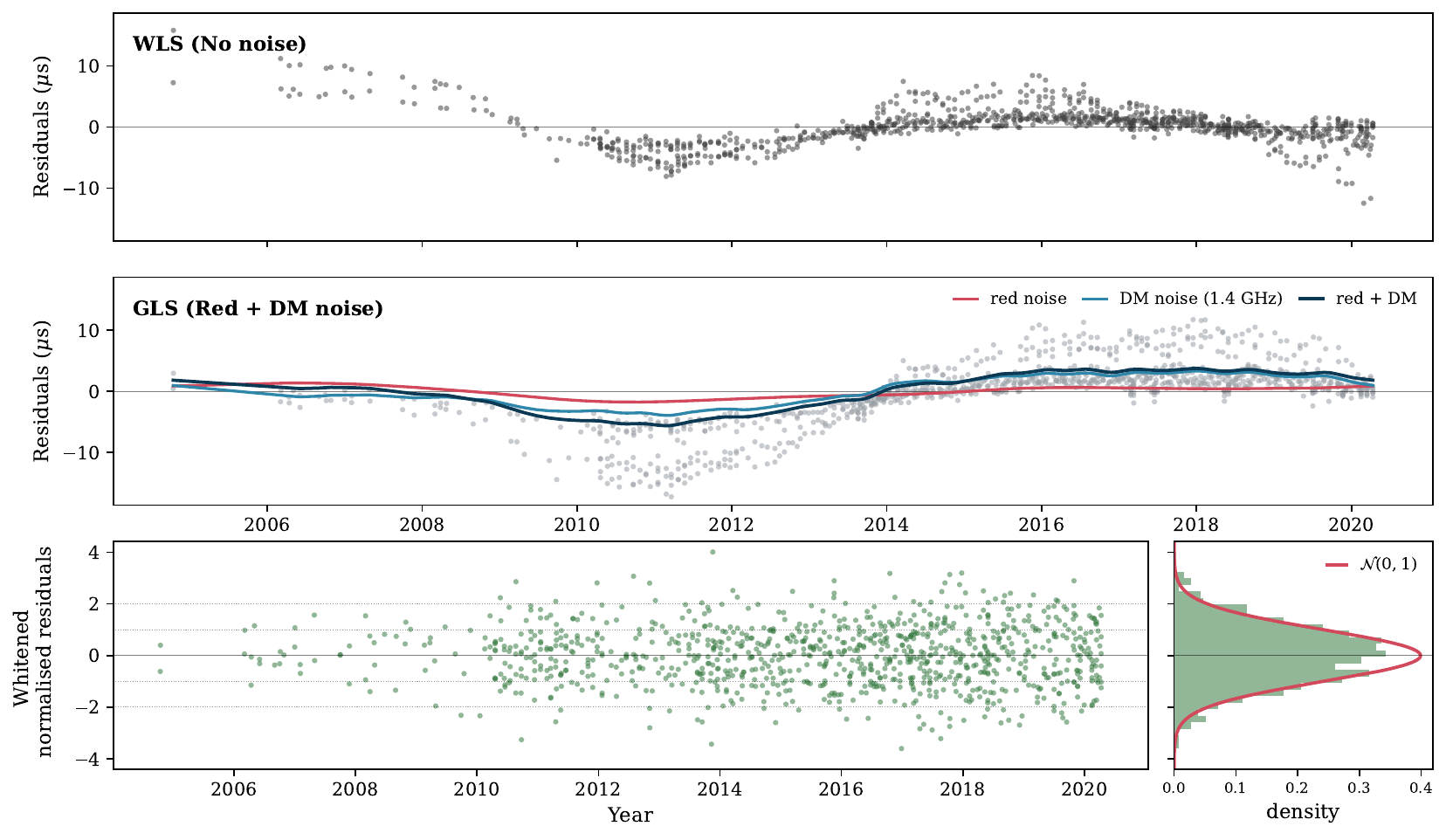}
    \caption{An example of WLS and GLS fitting using a simulated dataset of PSR~J1909$-$3744 containing achromatic red noise and DM noise. (Top) The post-fit residuals (grey) of a WLS solution, not modeling the injected noise processes. (Centre) The post-fit residuals (grey) of a GLS solution, taking into account the noise models. Overlaid on the residuals are the achromatic red noise (red) and DM noise (blue) as realised by \textsc{JUG} and referenced to $1.4\,\mathrm{GHz}$.  The curve in dark blue represents the sum of these processes. The visible differences between the noise realisations and the timing residuals are a product of the DM noise structure being referenced to a single radio frequency. (Bottom) The noise subtracted residuals normalised by their scaled uncertainties (green), demonstrated to be well approximated by a Gaussian distribution with mean $0$ and unit variance as expected.}
    \label{fig: J1909_wls_vs_gls}
\end{figure}

Section \ref{subsec: timing likelihood} established the penalised objective (\autoref{eq: penalised objective}) and that it is solved as a single augmented linear system, rather than by forming the large covariance matrix $\vect{C}$. \textsc{JUG} solves this system through a singular-value decomposition (SVD) where, in the WLS case, $\vect{\varepsilon}$ is found from the SVD of the whitened design matrix: $\vect{N}^{-1/2}\vect{M}$. In the GLS case, the timing and the noise bases are stacked into an augmented matrix $[\vect{M},\vect{F}]$ and recovered from an SVD of the same form. This SVD is used instead of forming the poorly conditioned matrix  $\vect{M}^\mathrm{T}\vect{N}^{-1}\vect{M}$. The fit iterates as described in Section \ref{subsec: timing likelihood}, re-evaluating $\vect{r}$ and $\vect{M}$ at the updated solution until convergence. \autoref{fig: J1909_wls_vs_gls} demonstrates the differences in fitting that occur from the use of each fitter on a noisy dataset.

\subsection{Noise Parameter Estimation}
\label{subsec: noise estimation}
The GLS fit described in Section \ref{subsec: fitting} relies on a fixed noise model, both white noise parameters and the hyperparameters of a power spectrum (amplitude and spectral index) are supplied, and the fit estimates the solution to the timing parameters given the noise. These noise parameters must also be determined by the data. Conventionally, this is done in a cyclic manner: an initial timing solution is formed, the noise parameters are determined from the resulting residuals, usually through sampling with the timing model marginalised, and the timing model parameters are measured again considering the noise.\footnote{This is true for workflows built from Enterprise \citep{2019ascl.soft12015E} and Discovery, however, we note that when using Temponest \citep{2014MNRAS.437.3004L} a post-noise ephemeris is produced.} To simplify this workflow, \textsc{JUG} enables a fast, in-package alternative to find the maximum-a-posteriori (MAP) point estimate of the noise hyperparameters. Point estimation of pulsar noise parameters from the marginalised likelihood is well established \citep[e.g.][]{2013MNRAS.428.1147V}, and \textsc{JUG}'s implementation is adapted directly from the \texttt{pulsar-map-noise-estimates}
package.\footnote{\url{https://github.com/davecwright3/pulsar-map-noise-estimates}}
\textsc{JUG} integrates this with the cached timing products and \textsc{JAX} machinery of the timing fit. The marginalised likelihood of the data given the noise model is implemented as a model using \textsc{Numpyro} \citep{2019arXiv191211554P}, and the MAP estimate is obtained through stochastic variational inference with a delta-function guide, optimised by gradient descent.\footnote{The algorithm uses an \textsc{Adam} optimiser \citep{kingma2014adam}.}

\begin{figure*}
    \centering
    \includegraphics[width=0.49\linewidth]{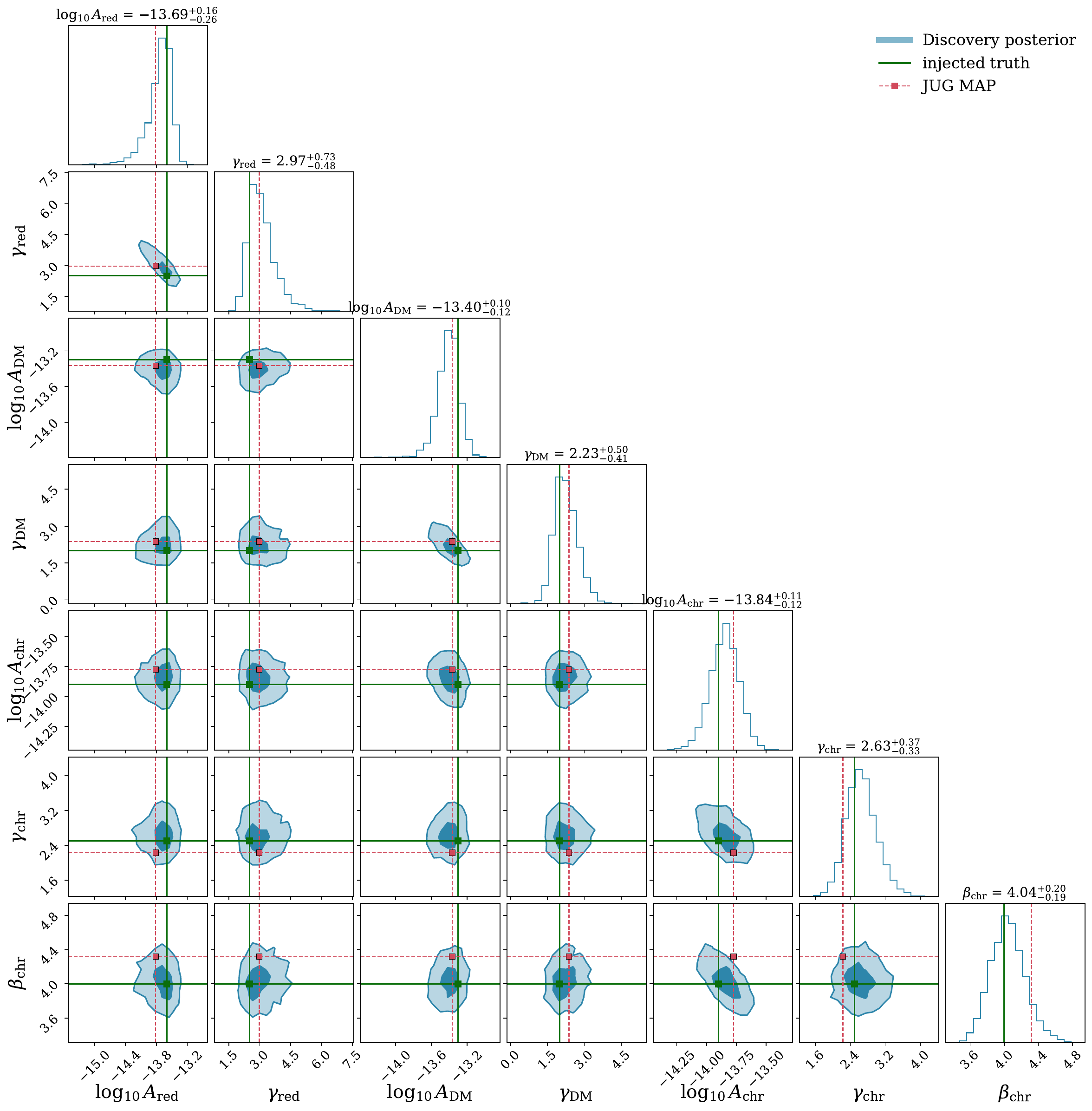}
    \hfill
    \includegraphics[width=0.49\linewidth]{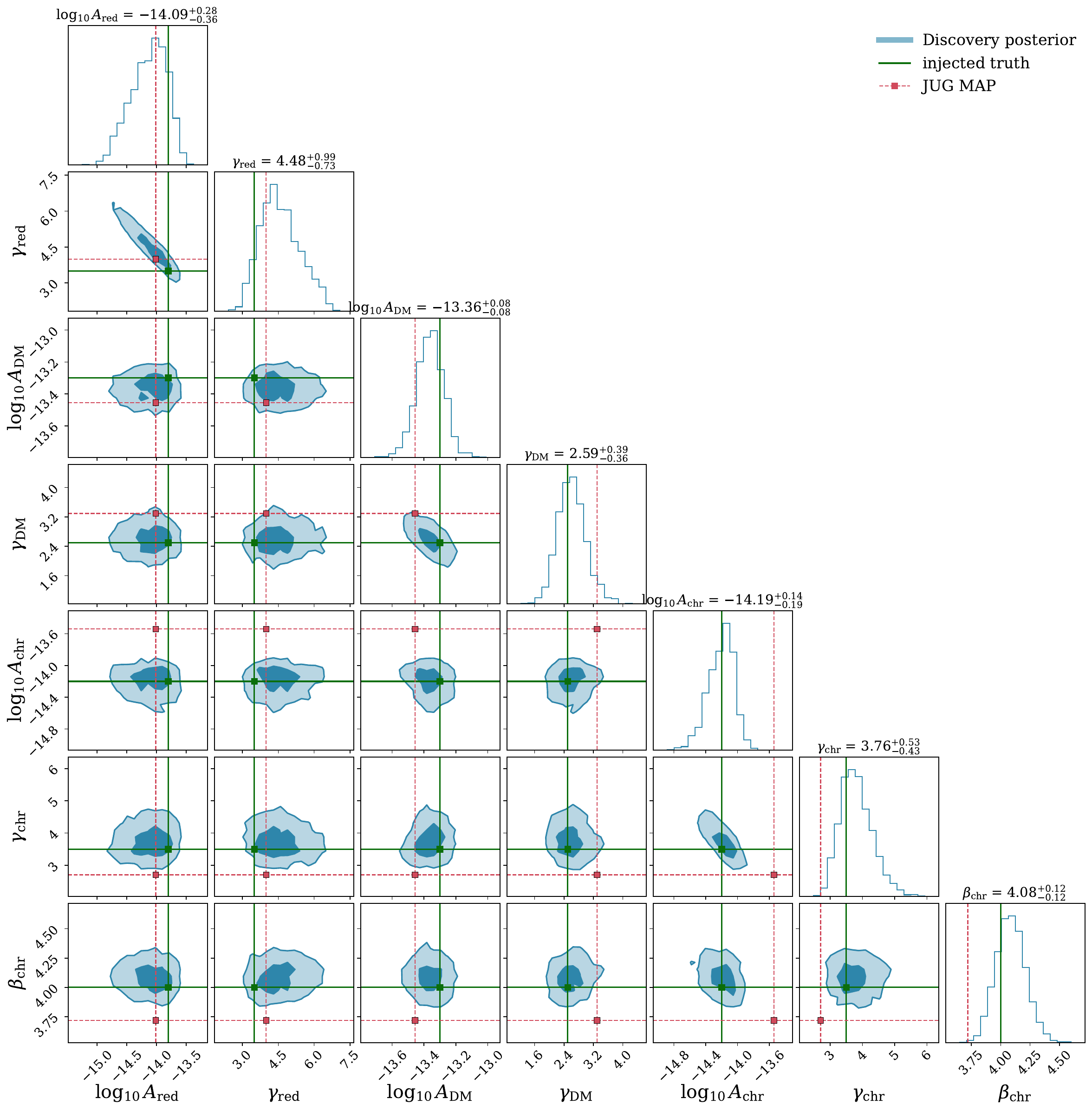}
    \caption{Comparison of \textsc{JUG}'s MAP noise estimates against posteriors sampled with \textsc{Discovery} for two simulated PSR~J1909$-$3744-like datasets with injected red, DM, and chromatic-scattering noise. Each parameter sampled here describes a hyperparameter of those noise processes. (Left) The MAP estimates (red) lie within the posteriors and agree with the injected values (green). (Right) The recovery of a dataset with steeper noise. The MAP recovery is mainly biased with reference to the chromatic amplitude. These illustrate the bias incurred where the noise processes are steep.}
    \label{fig: map_vs_discovery}
\end{figure*}

\textsc{JUG} estimates the white-noise parameters, achromatic red-noise, DM noise, and chromatic scattering noise with a variable chromaticity that can likewise be fit. Each of these time-correlated processes are modeled as a power-law spectrum with a Fourier basis. The MAP estimate returns a point estimate rather than a posterior so it is considerably faster than a thorough Bayesian analysis, however, the uncertainty information and the rigor of a Bayesian assessment is lost. Therefore it is best to consider the use of this module as a bridge between the frequentist timing regime and the full Bayesian treatment that usually follows. The accuracy of the point estimate depends on how well-constrained the noise processes are. For instance, for well-constrained spectra it can reach parity with a Bayesian analysis, however steep spectra that dominate the residuals can bias the estimate. \autoref{fig: map_vs_discovery} compares the MAP estimate framework in \textsc{JUG} against posteriors sampled with \textsc{Discovery}\footnote{\url{https://github.com/nanograv/discovery/tree/main}} for two simulated cases. The left demonstrates a well recovered spectrum that is shallower than the right, which biases the recovery of the parameters. Nevertheless, the MAP framework is a fast and well-motivated process that reduces potential bias in the initial fit timing solution, while providing a cross-check for a subsequent analysis.

The MAP estimates are obtained through iterating a variational optimisation thousands of times, a process which is very well suited to GPU acceleration. For warm evaluations, we find that it is approximately an order of magnitude faster on a GPU than a CPU, with the speed advantage growing as the dataset size increases. As a comparison, a $5000$ iteration optimisation on $50,000$ arrival times completes in $\sim19$ seconds on a GPU (NVIDIA GeForce RTX 4090) as compared to $\sim245$ seconds on a CPU (Section~\ref{sec: performance}).

\subsection{Data handling} 
\label{subsec: data handling}
\textsc{JUG} has its own parser and writer to handle ephemeris and timing files following the conventions of \textsc{Tempo2} \citep{2006MNRAS.369..655H, 2006MNRAS.372.1549E}, without dependencies on \textsc{Tempo2} or \textsc{PINT}\footnote{\textsc{JUG} reads \textsc{Tempo2}/IPTA format arrival time files, and is able to understand \texttt{MODE}, \texttt{TIME}, \texttt{INCLUDE}, and other comment directives.}. This independence is general to the full timing pipeline, with all the parsing, clock corrections, and barycentric geometry data computed within \textsc{JUG} itself (Section \ref{sec: Background on timing}) rather than wrapping or using other software. To support the pipeline without reliance on outside timing software, \textsc{JUG} ships with the additional data that the timing model requires, including the clock-correction files, solar system ephemeris, and observatory coordinates. Upon loading a dataset, \textsc{JUG} reads the arrival times and only parses the clock corrections for the observatories that are present in the data, rather than loading the correction chain for all known observatories. Several of these products are updated with some frequency, so \textsc{JUG} provides tools to keep them updated. Earth-orientation parameters and solar-system ephemerides are retrieved through \textsc{Astropy}, and the clock correction files are refreshed against the IPTA clock repository\footnote{\url{https://github.com/ipta/pulsar-clock-corrections}}, updating files that have become out of date. 

Upon reading a dataset, \textsc{JUG} validates the TOAs by identifying non-finite arrival times, frequencies, uncertainties, duplicated arrival times, or entries lacking system flags. For inspection of a timing solution, \textsc{JUG} can represent epoch-averaged residuals. The averaging is performed on the computed residuals rather than arrival times themselves as the residuals are formed in the high-precision pipeline and are already corrected for deterministic delays. An inverse-variance weighted average of the residuals within an epoch preserves the timing precision that would be lost when averaging the arrival times instead. \textsc{JUG} establishes the integer rotation count by phase-connected tracking: the first arrival time is rounded to the nearest turn and each subsequent count is advanced by the rounded phase difference from the previous arrival time, equivalent to \textsc{Tempo2}'s \texttt{TRACK -2} behaviour. This coincides with \autoref{eq: timing residual} whenever the timing solution holds all residuals within half a turn. Pulse numbers computed this way can be written to a pulse-numbered \texttt{.tim} file from the GUI or python API.

\subsection{Interfaces: Python API, CLI, and GUI} 
\label{subsec: interfaces}
\textsc{JUG} can be operated through three entry points. These are a set of command-line tools for quick checks, the Python API, and a graphical user interface (GUI), which all share the same computational engine and will therefore produce identical results (Section \ref{subsec: architecture}). The command-line tools cover the most common operations: \texttt{jug-compute-residuals} computes the timing residuals for a given timing model and associated arrival times, \texttt{jug-fit} creates a post-fit solution given the fitting flags present in the ephemeris file, and \texttt{jug-gui} launches the GUI. 

\begin{figure}
    \centering
    \includegraphics[width=\linewidth]{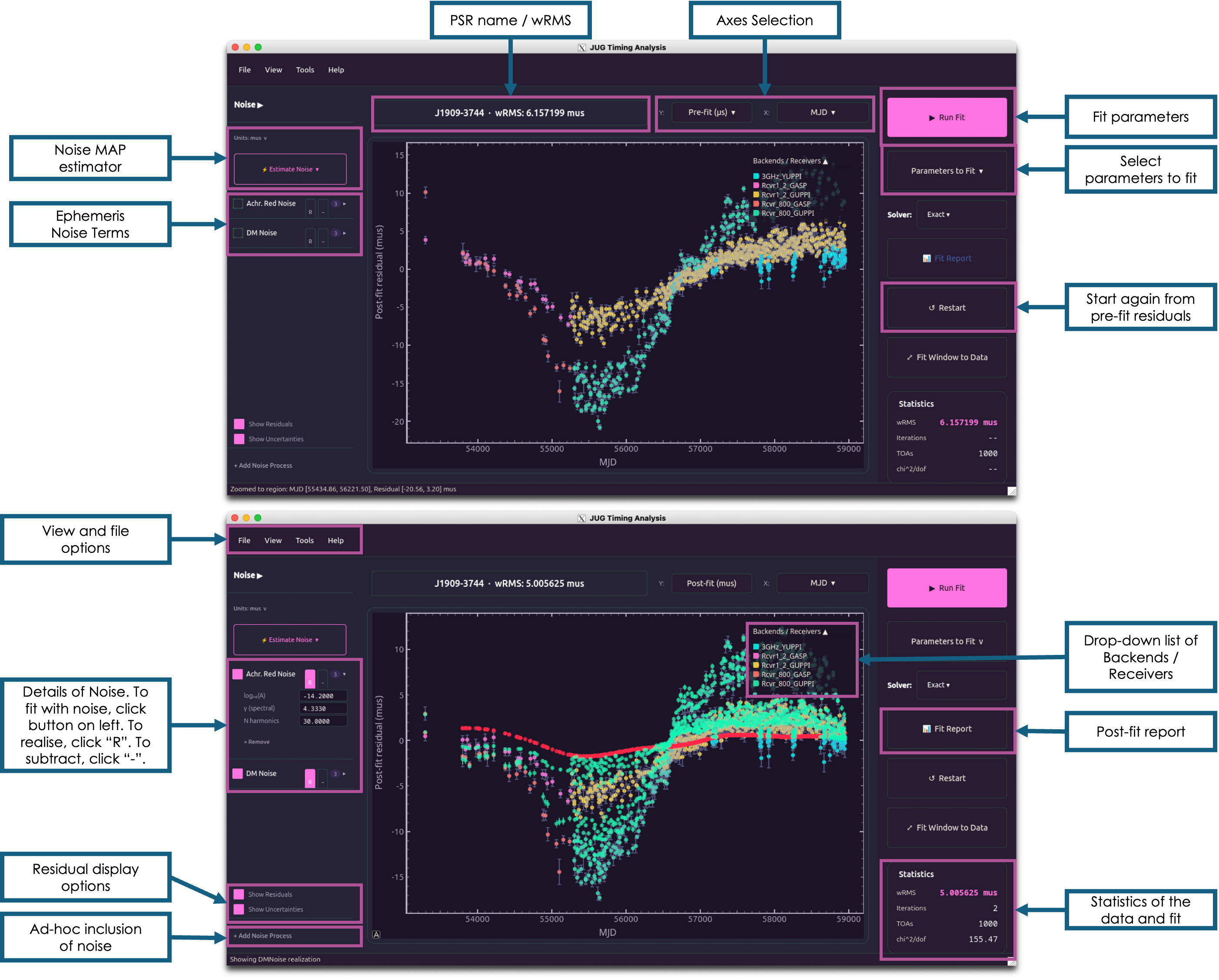}
    \caption{Examples of the JUG GUI on a simulated dataset of PSR~J1909$-$3744, demonstrating a loaded dataset's pre-fit residuals (top), and a post-fit solution with noise realisations overplotted (bottom). The figure is annotated to describe features of the GUI.}
    \label{fig: JUG GUI}
\end{figure}

The Python API is organised around the \texttt{TimingSession} object and was introduced earlier in this Section. From the Python API both the fitting pathways and the MAP estimator can be reached. An example of the code can be found in Appendix \ref{app: code}. The GUI is built to enable users to perform timing interactively. This is a common method of interacting with \textsc{Tempo2}, so we have implemented this to support the familiar workflow of those users coming from \textsc{Tempo2}. Using the GUI, a dataset can be loaded and fit interactively with a choice of which parameters or noise to include. In addition, residuals can be inspected across different axes including Modified Julian Date (MJD), orbital phase, or serial. An example of the GUI can be viewed in \autoref{fig: JUG GUI}.

\section{Performance}
\label{sec: performance}
We benchmark \textsc{JUG} against \textsc{PINT} and \textsc{Tempo2} on a controlled series of datasets mimicking and extending the NANOGrav 15-year PSR~J1909$-$3744 data, with the DMX entries replaced with a Taylor series expansion of DM. In this, we measure the computational cost of the fitting as a function of $N_\mathrm{TOA}$ and noise model complexity included in the ephemeris. These tests assess speed and scaling only, and the accuracy of the solutions will be considered in Section \ref{sec: validation}.

\subsection{Benchmarking Setup} 
Subsets of $N_\mathrm{TOA}$ are drawn, ranging from $250$ to $10^6$ arrival times. The data are injected with achromatic red noise and DM noise. We have considered two regimes in this benchmark: one which operates a WLS fit and another operating a GLS fit where achromatic red and DM noise are modelled with $60$ Fourier modes each. In both cases, \texttt{F0}, \texttt{F1}, \texttt{DM}, \texttt{DM1}, and \texttt{DM2} are fit for five iterations\footnote{The residual weighted RMS is stable and equivalent between software at all iterations past this number of iterations.}. All three software packages are tested on the same machine as their performance will be dependent on the underlying hardware: a workstation with an Intel Core Ultra $9$ $285$K CPU ($24$ cores), $64\,$GB of RAM, and an NVIDIA GeForce RTX $4090$ GPU ($24\,$GB). We note that the \textsc{Vela} software package is deliberately excluded from this comparison: \textsc{Vela} directly samples the timing model and noise parameters rather than solving a linearised system, so its computational cost is that of a stochastic sampler outputting a posterior rather than a least-squares fit producing a point estimate. Attempting to provide a computational comparison here would not be able to be done using like quantities.

\subsection{Benchmarking Results} 
\begin{figure}
    \centering
    \includegraphics[width=\linewidth]{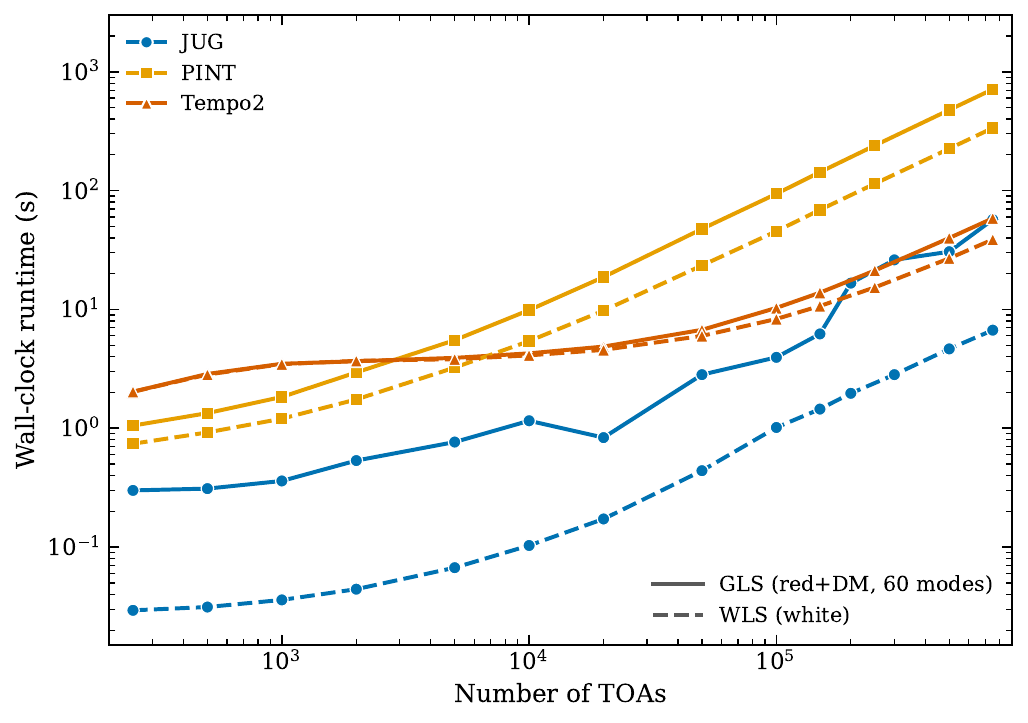}
    \caption{A computational speed comparison of JUG (blue), PINT (yellow), and Tempo2 (orange) evaluated through $N_\mathrm{TOA}$ using a simulated dataset for both GLS (solid) and WLS (dashed) fits. While \textsc{JUG} is the fastest across the range of dataset sizes tested, the \textsc{Tempo2} \texttt{-qrfit} solver approaches \textsc{JUG}'s speed as the data size increases and noise is included in the pulsar ephemeris.}
    \label{fig: scaling figure}
\end{figure}

\begin{figure}
    \centering
    \includegraphics[width=\linewidth]{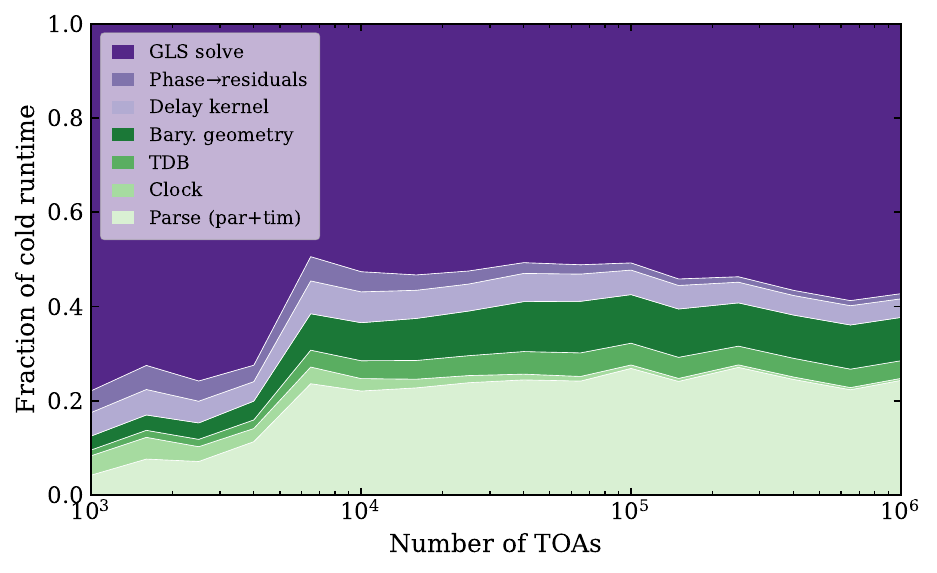}
    \caption{The fractional breakdown of \textsc{JUG}'s computation for a fresh load and fit versus the number of arrival times using the GLS fitter. The fractional share of the GLS computation drops at $\sim6\times10^3$ arrival times, once the stages that scale with arrival times begin to dominate.}
    \label{fig: JUG_breakdown}
\end{figure}

\autoref{fig: scaling figure} shows the wall-clock runtime as a function of $N_\mathrm{TOA}$ for each package. In the WLS case, \textsc{JUG} is faster than \textsc{PINT} by a factor growing from $\sim25\times$ at $250$ TOAs to $\sim56\times$ at $20{,}000$ TOAs, with the relative speed of \textsc{JUG} growing as the dataset sizes increase. In the GLS regime, where the correlated noise model increases the number of design matrix columns, \textsc{JUG} is faster than \textsc{PINT} by factors\footnote{This is heavily dependent on the dataset and noise model assessed. The largest increase in speed for \textsc{JUG} over \textsc{PINT} is $>50\times$.} of $\sim5-20\times$. Against \textsc{Tempo2}'s \texttt{-qrfit} solver, \textsc{JUG} is several times faster on the scale of small to intermediate sized datasets, but become comparable at larger datasets\footnote{We note that Tempo2's default SVD solver demonstrates approximately the same speed as PINT.}, converging to similar runtimes as $N_\mathrm{TOA}$s approaches $10^6$ (\autoref{fig: scaling figure}).
The per-TOA computations and linear algebra are written to run on either a CPU or a GPU via \textsc{JAX}. We tested the performance of both capabilities, initialising the data from a cold-start and repeating fits so that the caching strategy activates, comparing only the linear solve. For these tests the GPU offered no advantage over the CPU for the timing fit, demonstrating that at these dataset sizes GPU acceleration of the linear solve is not favourable. The noise estimation outlined in Section \ref{subsec: noise estimation}, however, does benefit substantially from the use of a GPU. For this, the GPU pathway is a useful feature to have in \textsc{JUG}'s design. Future, substantially larger, datasets will appear with more complex noise models than we implement here, and at these scales the per-TOA computations may begin to outweigh the compilation and host-to-GPU transfer costs. \autoref{fig: JUG_breakdown} decomposes \textsc{JUG} runtime into the stages of its pipeline as a function of $N_\mathrm{TOA}$, separating the one-time operations that are subsequently cached across the session (Section \ref{subsec: architecture}) from those evaluated at every iteration. At small dataset sizes, the linear solve dominates the runtime, as its cost is set by the number of design-matrix columns which is independent of $N_\mathrm{TOA}$. As the dataset increases, the stages that scale with $N_\mathrm{TOA}$ begin to dominate and the fractional cost of the solve falls for the runtime. This demonstrates that even with the optimisation that \textsc{JUG} employs in its solve, it is still the single largest computational stage and is naturally the target for future improvements. 

\section{Validation}
\label{sec: validation}

The performance gains described in Section \ref{sec: performance} are only worthwhile should \textsc{JUG} produce correct results. A simple way to demonstrate this is validating \textsc{JUG} by confirming that it reproduces the solutions of \textsc{PINT} at the precision that is required for pulsar timing, and that it is able to recover known parameters in simulated data.

\subsection{Residual Level Agreement} 
\label{subsec: residual agreement}
\begin{figure}
    \centering
    \includegraphics[width=\linewidth]{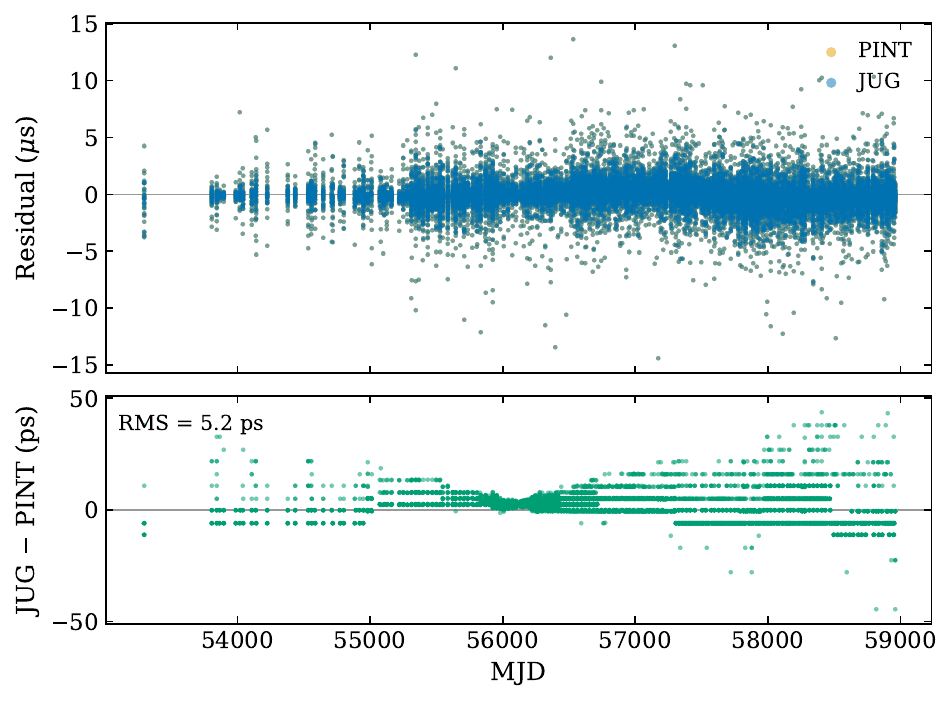}
    \caption{(Top) The residuals of \textsc{JUG} and \textsc{PINT} overlaid on top of each other validated at the same timing model parameter values for PSR~J1909$-$3744 from the NANOGrav 15-year data set. It is difficult to see the \textsc{PINT} data as \textsc{JUG} matches them almost exactly. (Bottom) The residual difference between the residuals formed by each software, demonstrating ps level of agreement and a relative RMS difference between them of $5.25\,\mathrm{ps}$. The bowtie structure seen in the bottom panel is a signature of the dataset growth from the \texttt{PEPOCH} set in the ephemeris. The horizontal banding reflects the quantization due to precision handling of the phase residual.}
    \label{fig: JUG_PINT_validation}
\end{figure}

We evaluate the residuals of PSR~J1909$-$3744 from the NANOGrav 15-year dataset \citep{2023ApJ...951L...9A} in both \textsc{JUG} and \textsc{PINT} at the same given parameters, using the same solar-system ephemeris (DE440), timescale (TDB) and clock corrections. \autoref{fig: JUG_PINT_validation} shows this result, demonstrating that the residuals between the two packages are virtually indistinguishable. The RMS of the difference between them is $5.25\,\mathrm{ps}$ over the $15$ year baseline, more than four orders of magnitude below the uncertainties on the arrival times. This agreement is distinct from \textsc{JUG}'s precision against exact arithmetic (Section \ref{subsec: precision}). \textsc{PINT} and \textsc{JUG} share a strategy to account for the extended precision of the phase, so they track each other more closely than either tracks the exact arithmetic. The difference is smallest at the centre of the dataset where \texttt{PEPOCH} is defined in the pulsar ephemeris, growing to either end. This is the signature of the accumulation of the precision representations between \textsc{JUG} and \textsc{PINT}, a fundamental precision floor rather than any disagreement in the timing model.
\subsection{Fit Level Parity} 
\label{subsec: fit parity}

In Section \ref{subsec: residual agreement} we established that \textsc{JUG} and \textsc{PINT} compute the same residuals at given identical parameters. We now confirm that they converge to the same solution through fitting. To test this we take two pulsars from the NANOGrav 15-year dataset: PSR~J1909$-$3744 and PSR~J0437$-$4715. PSR~J1909$-$3744 is in a \texttt{ELL1} binary and is generally considered one of the best millisecond pulsars to time, and PSR~J0437$-$4715 is the closest and brightest millisecond pulsar with a \texttt{DD} binary model. Both are fit in \textsc{JUG} and \textsc{PINT} using the same data and the same fitted parameters, with the ephemeris, clock corrections and timescales identical, using a GLS fit to include the noise model.

We assess the parity of these fits by comparing each post-fit parameter against the difference expected from its reported parameter uncertainty, such that for a parameter $p$ with uncertainty $\sigma_p$, we report $|p_\mathrm{JUG} - p_\mathrm{PINT}|/\sigma_p$. A summary of the results of this are reported in Table \ref{tab: parity_summary} with the full results displayed in Table \ref{tab: app_jug_pint_fit_parity}. The table here shows that between the two software packages, the fitted parameters agree to better than $0.014\sigma$, and no parameter difference exceeds $0.1\sigma$. This shows that \textsc{JUG} is able to reproduce \textsc{PINT}'s timing solution to a small fraction of the parameter uncertainties.

\begin{table}[]
    \centering
    \begin{tabular}{l c c c c c}
        \hline\hline
        Pulsar & $N_\mathrm{TOA}$ & Binary & RMS ($\mu$s) & $\max|\Delta|/\sigma$ & $N_{>0.1\sigma}$ \\
        \hline
        J1909$-$3744 & $35{,}037$ & ELL1 & $0.365$ & $0.0136$ & $0$ \\
        J0437$-$4715 & $5{,}830$  & DD   & $0.398$ & $0.0072$ & $0$ \\
        \hline\hline
    \end{tabular}
    \caption{Summary of the fit parity on two NANOGrav 15-year pulsars, fit with GLS including the noise model. For both pulsars we list the number of TOAs, binary model, post-fit weighted RMS, the maximum per-parameter difference in units of the formal uncertainty, and the number of parameters differing by more than $0.1\sigma$. The full per-parameter comparison is given in Appendix~\ref{app: parity}.}
    \label{tab: parity_summary}
\end{table}

The parity above holds as these fits are well-conditioned and the data well-described by the model. Where this is not the case, the objective may become degenerate in the covariant directions. In a valley such as this, the data are not able to constrain the parameter combinations consistently across any timing software, amplifying the inherent differences between the software into large parameter differences. A fit on a deliberately degenerate pulsar, the same PSR~J1909$-$3744 but without the noise included in the model, demonstrates that \textsc{JUG}, \textsc{PINT}, and \textsc{Tempo2} all diverge from each other on the order of tens of $\sigma$ in the covariant parameters, with no individual software the outlier. This divergence is a reflection of the conditioning of the problem, rather than in the timing machinery of any of the programs.

\subsection{Injection Recovery} 
\label{subsec: injection recovery}
The agreement between \textsc{JUG} and \textsc{PINT} in Sections \ref{subsec: residual agreement} and \ref{subsec: fit parity} establishes that \textsc{JUG} is consistent with \textsc{PINT} by design. However this does not necessarily establish that either software is able to recover the true parameters of a dataset. To test this directly, we construct a simulated dataset to verify the ability of \textsc{JUG} to converge to a known truth that is simulated independently of \textsc{JUG}. We create this simulated dataset by mimicking PSR~J1600$-$3053 from the NANOGrav 15 year dataset, chosen as it is a well understood binary. We alter this pulsar's model slightly by removing the \texttt{DMX} parameters and replacing them with a smooth Taylor-series approximation of the DM, and then establish a ground truth ephemeris and adjust the arrival times using \textsc{PINT} so that the residuals are zero when represented by these chosen parameters. Following this, we perturb each of the fitted parameters by $\pm3\sigma$ of each parameter's uncertainty and fit from this point using both packages, recovering the injected values. 

\begin{figure}
    \centering
    \includegraphics[width=\linewidth]{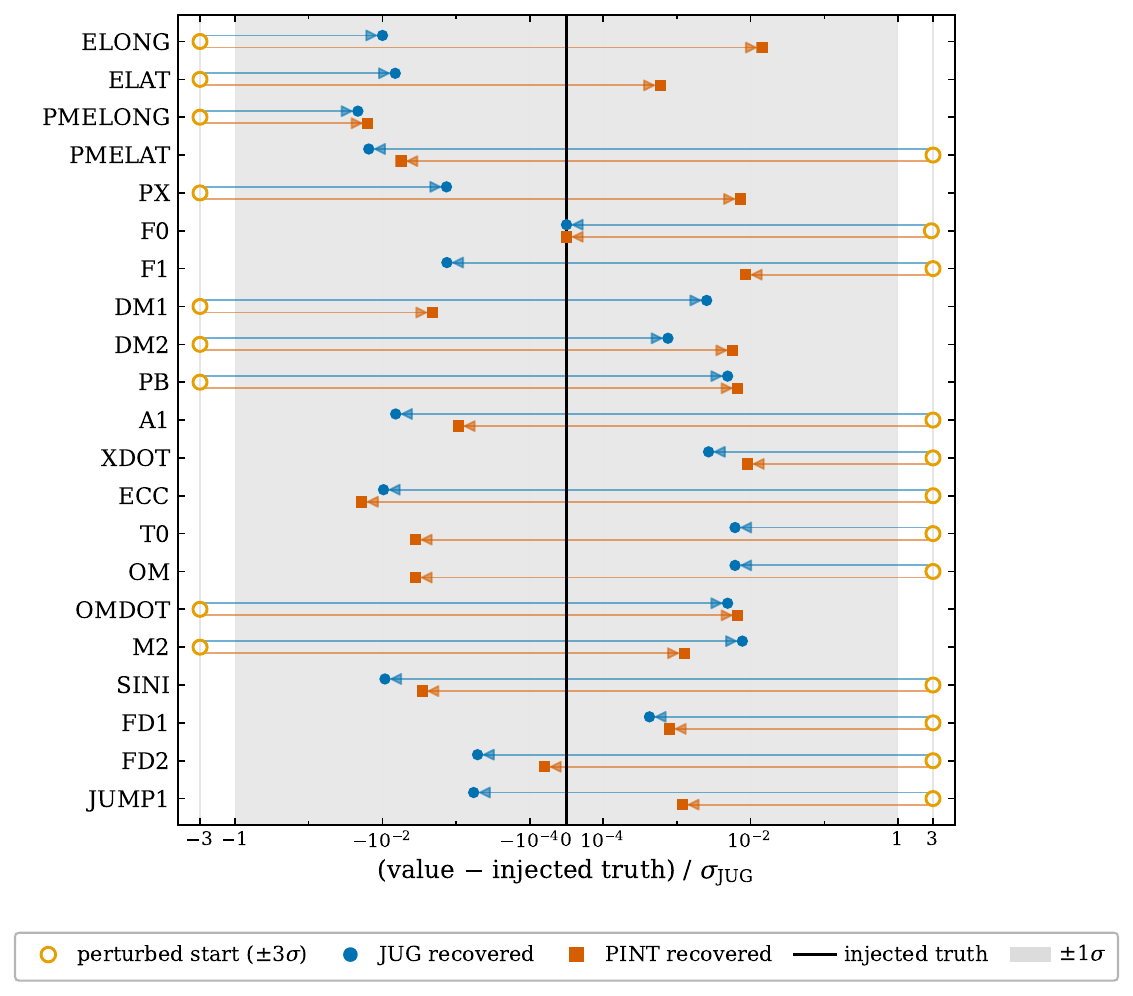}
    \caption{The recovery of the timing parameters of a simulated dataset of PSR~J1600$-$3053 for both \textsc{JUG} (blue) and \textsc{PINT} (orange), from a starting ephemeris where all parameters were perturbed by $\pm3\sigma$ from their simulated values. Both \textsc{JUG} and \textsc{PINT} recover the timing solution to approximately the same accuracy, within $0.02\sigma$ of each parameter's true simulated value. The greyed out region in the centre of the plot represents the $\pm1\sigma$ region of parameter uncertainty.}
    \label{fig: JUG_PINT_inj_recovery}
\end{figure}

\autoref{fig: JUG_PINT_inj_recovery} shows this result, demonstrating that both packages are able to recover each parameter to far below the formal uncertainties of the parameters. Neither code is systematically closer to the truth, with each recovering some parameters marginally better than the other. From this we conclude that \textsc{JUG} is able to recover the true parameters of an independently constructed dataset to the same level as \textsc{PINT}.  

\section{Summary and Future Work}
\label{sec: summary}
\textsc{JUG} is a \textsc{JAX} based pulsar timing package, built to operate independently of other available timing software while possessing demonstrable parity with existing packages. It is comparably fast to the compiled \textsc{Tempo2} and at least an order of magnitude faster than \textsc{PINT}. It has been validated to agree with \textsc{PINT} to the picosecond level while not relying on any architecture found in either \textsc{PINT} or \textsc{Tempo2}. It is designed to be operated through multiple entry points, combining fitting, interactive use, and noise estimation within a single fast and independent package. \textsc{JUG} does require a one-time JIT compilation cost, however, even accounting for this the speed is still equivalent to or greater than the otherwise fastest timing software. Further, it possesses full GPU capabilities. While GPU acceleration is not yet advantageous for the timing fit at current dataset sizes, the MAP noise estimation completes up to an order of magnitude faster on a GPU than a CPU. We did not compare \textsc{JUG} against the \textsc{Vela} timing package directly, as \textsc{JUG} and \textsc{Vela} approach the divide between frequentist and Bayesian analyses from opposite directions. \textsc{Vela} replaces the fitting process by sampling the timing model, while \textsc{JUG} accelerates the fit and bridges towards the Bayesian regime through a MAP noise estimation. The two can be considered complementary rather than competitive approaches.

As \textsc{JUG} currently stands, it is a fully capable timing package verified against pulsars from the NANOGrav $15$-year dataset. Naturally, this testing data does not amount to all pulsar timing data, nor the specific requirements other datasets may possess, so \textsc{JUG} is not yet expected to be a general purpose timing program. However it is being actively developed, and we anticipate that it will become a general purpose tool. In future work we plan to extend \textsc{JUG} to wideband timing, fully Bayesian timing, fitting deterministic signals via \textsc{JAX} auto-differentiation, and an extended GPU-based pipeline. Currently, \textsc{JUG} presumes a phase-connected timing solution and refines it. It does not yet provide a mechanism for the interactive insertion of pulse turns or automated solution finding in the style of \textsc{psrtime}, \textsc{Dracula} \citep{2018MNRAS.476.4794F} or the algorithmic approach presented in \citet{2024ApJ...964..128T}, which we consider a natural direction for future development. Further, we intend to integrate it with profile domain timing with the \textsc{SCULPT} package (Miles et al., in prep.), and the in-development PTA global fit software package \textsc{ATLAS} (Gersbach et al., Laal et al., in prep.). In support of this, \textsc{JUG} is open-source and welcomes contributions. Interested users are encouraged to contact the corresponding author.

\begin{acknowledgments}
We acknowledge the use of large language model based tools during the development of the \textsc{JUG} codebase. The tools assisted with code implementation and all resulting code was reviewed, tested and validated by the authors. The correctness of \textsc{JUG}'s outputs is detailed in Section \ref{sec: validation}. MM, SRT, and NL acknowledge support from the NANOGrav Collaboration's National Science Foundation (NSF) Physics Frontiers Center award numbers 1430284, 2020265, and 2607948. Parts of this work were supported by the Australian Research Council Centre of Excellence of Gravitational Wave Discovery (CE230100016). SRT and KAG acknowledge support from NSF CAREER \#2146016. SRT also acknowledges support from NSF AST-2307719, NSF NRT-2125764, NASA LPS-80NSSC26K0342, and a Chancellor's Faculty Fellowship from Vanderbilt University. The National Radio Astronomy Observatory is a facility of the National Science Foundation operated under cooperative agreement by Associated Universities, Inc. SMR is a CIFAR Fellow and is supported by the NSF Physics Frontiers Center award 2020265 and NSF AAG award 2510064. ML acknowledges support from NSF Physics Frontiers Center award number 2020265 and 2607948 (NANOGrav), NSF AAG award number 2511107. N.L. is supported by the Vanderbilt Initiative in Data Intensive Astrophysics (VIDA) Fellowship. This work was conducted in part using the resources of the Advanced Computing Center for Research and Education (ACCRE) at Vanderbilt University, Nashville, TN. A.P acknowledge financial support from the European Research Council (ERC) starting grant 'GIGA' (grant agreement number: 101116134). A.P. also acknowledges support through the NWO-I Veni fellowship. We acknowledge financial support from the "Programme National de Cosmologie et Galaxies" (PNCG), "Programme National Hautes Energies" (PNHE) and "Programme National Gravitation, Références, Astronomie, Métrologie" (PNGRAM) of CNRS/INSU, France. We acknowledge financial support from Agence Nationale de la Recherche (ANR-18-CE31-0015), France. 
\end{acknowledgments}

\software{
JAX \citep{jax2018github},
NumPy \citep{2020Natur.585..357H},
SciPy \citep{2020NaMet..17..261V},
Astropy \citep{2022ApJ...935..167A},
ERFA,
NumPyro \citep{2019arXiv191211554P},
optax \citep{deepmind2020jax},
PySide6,
PINT \citep{2021ApJ...911...45L},
Tempo2 \citep{2006MNRAS.369..655H}
}

\appendix

\section{Parameter Support Tables}
\label{app: parameters}
Tables \ref{tab: binary_models} and \ref{tab: components} summarise the binary timing models and the deterministic and stochastic model components implemented in \textsc{JUG}.

\begin{table*}[h]
\centering
\arrayrulecolor{gray!40}
\begin{tabular}{ll@{\hspace{1.5cm}}l}
\arrayrulecolor{black}\hline\hline
Model & Description & \makecell[l]{Principal fitted parameters} \\
\arrayrulecolor{black}\hline
\texttt{BT} & Blandford--Teukolsky \citep{1976ApJ...205..580B} & \makecell[l]{PB, A1, ECC, OM, T0, OMDOT,\\ EDOT, GAMMA, PBDOT} \\
\arrayrulecolor{gray!40}\hline
\texttt{DD} & Damour--Deruelle \citep{1986AIHPA..44..263D} & \makecell[l]{PB, A1, ECC, OM, T0, OMDOT, EDOT,\\ PBDOT, GAMMA, M2, SINI} \\
\arrayrulecolor{gray!40}\hline
\texttt{DDS} & DD with alternative Shapiro shape & \makecell[l]{DD parameters with\\ SHAPMAX replacing SINI} \\
\arrayrulecolor{gray!40}\hline
\texttt{DDH} & DD with orthometric Shapiro & \makecell[l]{DD parameters with\\ H3, H4, STIG} \\
\arrayrulecolor{gray!40}\hline
\texttt{DDGR} & DD in general relativity & \makecell[l]{PB, A1, ECC, OM, T0, MTOT, M2\\ (post-Keplerian terms derived\\ from the masses)} \\
\arrayrulecolor{gray!40}\hline
\texttt{DDK} & DD with Kopeikin terms \citep{1995ApJ...439L...5K, 1996ApJ...467L..93K} & \makecell[l]{DD parameters with\\ KIN, KOM} \\
\arrayrulecolor{gray!40}\hline
\texttt{ELL1} & Low-eccentricity \citep{2001MNRAS.326..274L} & \makecell[l]{PB, A1, TASC, EPS1, EPS2,\\ EPS1DOT, EPS2DOT, PBDOT, M2, SINI} \\
\arrayrulecolor{gray!40}\hline
\texttt{ELL1H} & ELL1 with orthometric Shapiro & \makecell[l]{ELL1 parameters with\\ H3, H4, STIG} \\
\arrayrulecolor{gray!40}\hline
\texttt{T2} & General model, dispatched as below \citep{2006MNRAS.369..655H} & \makecell[l]{Inferred from the ephemeris\\ parameters present} \\
\arrayrulecolor{black}\hline\hline
\end{tabular}
\arrayrulecolor{black}
\caption{Binary timing models supported by \textsc{JUG} and the principal parameters fitted by each. Derivatives are analytic for all models except \texttt{BT}, which uses automatic differentiation.}
\label{tab: binary_models}
\end{table*}

\begin{table}[h]
\centering
\arrayrulecolor{black}
\begin{tabular}{ll}
\arrayrulecolor{black}\hline\hline
Component & \makecell[l]{Representative parameters} \\
\arrayrulecolor{black}\hline
Spin / phase & \makecell[l]{F0--F$n$, PEPOCH} \\
\arrayrulecolor{gray!40}\hline
Astrometry (equatorial) & \makecell[l]{RAJ, DECJ, PMRA, PMDEC} \\
\arrayrulecolor{gray!40}\hline
Astrometry (ecliptic) & \makecell[l]{ELONG, ELAT, PMELONG, PMELAT} \\
\arrayrulecolor{gray!40}\hline
Parallax & \makecell[l]{PX} \\
\arrayrulecolor{gray!40}\hline
Dispersion & \makecell[l]{DM, DM1--DM$n$, DMEPOCH, DMX} \\
\arrayrulecolor{gray!40}\hline
Solar wind & \makecell[l]{NE\_SW} \\
\arrayrulecolor{gray!40}\hline
Frequency-dependent & \makecell[l]{FD1--FD$n$, FD jumps} \\
\arrayrulecolor{gray!40}\hline
Glitches & \makecell[l]{GLEP, GLPH, GLF0, GLF1, GLF0D, GLTD} \\
\arrayrulecolor{gray!40}\hline
Chromatic events & \makecell[l]{exponential dips, chromatic\\ exponential and annual events} \\
\arrayrulecolor{gray!40}\hline
System offsets & \makecell[l]{phase jumps (JUMP)} \\
\arrayrulecolor{gray!40}\hline
Deterministic GW & \makecell[l]{continuous waves,\\ bursts with memory} \\
\arrayrulecolor{gray!40}\hline
Other delays & \makecell[l]{troposphere, planetary Shapiro delay} \\
\arrayrulecolor{gray!40}\hline
White noise & \makecell[l]{EFAC, EQUAD, ECORR} \\
\arrayrulecolor{gray!40}\hline
Correlated noise & \makecell[l]{power-law Gaussian processes:\\ achromatic red, DM, chromatic scattering} \\
\arrayrulecolor{black}\hline\hline
\end{tabular}
\arrayrulecolor{black}
\caption{Deterministic and stochastic model components implemented in \textsc{JUG}, with representative parameters for each.}
\label{tab: components}
\end{table}

\section{Code Examples \& Interface Reference Material}
\label{app: code}

\autoref{fig: api example} demonstrates the core \textsc{JUG} Python API workflow: loading a pulsar dataset, computing residuals, fitting the timing model with the noise model and free parameters read from the ephemeris, estimating the noise-model parameters, and writing the post-fit ephemeris. The starting ephemeris has had its fitted parameters deliberately perturbed away from the converged solution, so the example demonstrates the fit recovering the timing solution. \autoref{fig: api example interactive} demonstrates interactive control of the fitted parameter set: restricting the fit to a subset of parameters, and freeing or introducing parameters without editing the ephemeris file. The executed notebook versions of these examples are available in the \textsc{JUG} repository.

\begin{figure}[h]
\begin{verbatim}
>>> import os
>>> os.environ["JAX_PLATFORMS"] = "cpu"   # optionally switch to CPU driven code

>>> from jug.engine.session import TimingSession
>>> session = TimingSession("J1909-3744.par", "J1909-3744.tim")
>>> print(session)
TimingSession(par='J1909-3744.par', tim='J1909-3744.tim')

>>> pre = session.compute_residuals()
>>> print(f"pre-fit RMS = {pre['rms_us']:.3f} us")
pre-fit RMS = 101.771 us

>>> fit = session.fit_parameters()   # GLS fit; noise model and free
...                                  # parameters read from the ephemeris
>>> print(f"post-fit RMS = {fit['final_rms']:.3f} us, converged = {fit['converged']}")
post-fit RMS = 0.497 us, converged = True

>>> noise = session.estimate_noise()   # MAP estimate of the noise model
>>> print(f"log10 A_DM = {noise.params['TNDMAMP']:.2f}")
log10 A_DM = -13.47

>>> session.save_par("J1909-3744_postfit.par", fit_result=fit)
\end{verbatim}
\caption{Code example showing the core \textsc{JUG} API used to time PSR~J1909$-$3744. A \texttt{TimingSession} is constructed from the ephemeris and arrival-time files, residuals are computed, and the timing model is fit with a generalised least-squares solve using the noise model carried in the ephemeris. The starting ephemeris is deliberately perturbed, and the fit recovers the solution, reducing the residual RMS by over two orders of magnitude. The noise-model hyperparameters are then estimated with the maximum-a-posteriori framework of Section \ref{subsec: noise estimation}, and the post-fit ephemeris is written out.}
\label{fig: api example}
\end{figure}

\begin{figure}[h]
\begin{verbatim}
>>> sub = TimingSession("J1909-3744.par", "J1909-3744.tim")

>>> # Fit a subset: freeze everything except F0, F1, DM1
>>> sub.set_frozen([p for p in sub.free_params if p not in ("F0", "F1", "DM1")])
>>> print("fitting:", sub.free_params)
fitting: ['DM1', 'F0', 'F1']

>>> # Free a parameter the par file carries but does not flag (DM),
>>> # and one the par file omits entirely (F2, starting from zero)
>>> sub.set_free("DM", "F2")
>>> sub_fit = sub.fit_parameters()
>>> print(f"F2 = {sub_fit['final_params']['F2']:.3e}")
F2 = -1.290e-25
\end{verbatim}
\caption{Interactive control of the fitted parameter set through the \textsc{JUG} API. The fit flags are edited in memory with \texttt{set\_frozen()} and \texttt{set\_free()} rather than by modifying the ephemeris file: the fitted set is restricted to a chosen subset, a parameter present in the ephemeris but not flagged for fitting (\texttt{DM}) is freed, and a parameter absent from the ephemeris entirely (\texttt{F2}) is introduced through the parameter registry and included in the fit.}
\label{fig: api example interactive}
\end{figure}

\begin{table}[h]
\centering
\arrayrulecolor{black}
\begin{tabular}{ll}
\arrayrulecolor{black}\hline\hline
Key / interaction & \makecell[l]{Action} \\
\arrayrulecolor{black}\hline
\multicolumn{2}{l}{\textit{Files and session}} \\
\arrayrulecolor{gray!40}\hline
Ctrl+P & \makecell[l]{Open .par file} \\
\arrayrulecolor{gray!40}\hline
Ctrl+T & \makecell[l]{Open .tim file} \\
\arrayrulecolor{gray!40}\hline
Ctrl+S & \makecell[l]{Save .par file} \\
\arrayrulecolor{gray!40}\hline
Ctrl+Shift+S & \makecell[l]{Save .tim file} \\
\arrayrulecolor{gray!40}\hline
Ctrl+N & \makecell[l]{Save pulse-numbered .tim file} \\
\arrayrulecolor{gray!40}\hline
Ctrl+E & \makecell[l]{Open the parameters panel} \\
\arrayrulecolor{gray!40}\hline
Ctrl+Q & \makecell[l]{Exit} \\
\arrayrulecolor{black}\hline
\multicolumn{2}{l}{\textit{Fitting and TOAs}} \\
\arrayrulecolor{gray!40}\hline
Ctrl+F & \makecell[l]{Run the fit} \\
\arrayrulecolor{gray!40}\hline
Ctrl+R & \makecell[l]{Restart (reset to the original ephemeris)} \\
\arrayrulecolor{gray!40}\hline
Ctrl+A & \makecell[l]{Average TOAs (dialog)} \\
\arrayrulecolor{gray!40}\hline
Ctrl+Shift+A & \makecell[l]{Restore original TOAs} \\
\arrayrulecolor{black}\hline
\multicolumn{2}{l}{\textit{Plot interaction}} \\
\arrayrulecolor{gray!40}\hline
Z & \makecell[l]{Box zoom: first press sets a corner at the\\ cursor, second press applies the zoom} \\
\arrayrulecolor{gray!40}\hline
Shift+Z & \makecell[l]{Box delete: first press sets a corner,\\ second press deletes the TOAs in the box} \\
\arrayrulecolor{gray!40}\hline
Ctrl+0 & \makecell[l]{Zoom to fit} \\
\arrayrulecolor{gray!40}\hline
U & \makecell[l]{Unzoom to the full view} \\
\arrayrulecolor{gray!40}\hline
Esc & \makecell[l]{Cancel an active box zoom or box delete;\\ closes an open overlay menu first if present} \\
\arrayrulecolor{black}\hline
\multicolumn{2}{l}{\textit{Display}} \\
\arrayrulecolor{gray!40}\hline
W & \makecell[l]{Toggle weighted / unweighted RMS\\ in the statistics panel} \\
\arrayrulecolor{gray!40}\hline
R & \makecell[l]{Toggle residual error-bar visibility} \\
\arrayrulecolor{gray!40}\hline
Shift+K & \makecell[l]{Toggle subtraction of the red-noise realization} \\
\arrayrulecolor{gray!40}\hline
Shift+F & \makecell[l]{Toggle subtraction of the DM-noise realization} \\
\arrayrulecolor{black}\hline
\multicolumn{2}{l}{\textit{Mouse}} \\
\arrayrulecolor{gray!40}\hline
Left click, backend legend & \makecell[l]{Expand or collapse the backend\\ colour legend} \\
\arrayrulecolor{gray!40}\hline
Left click, noise panel row & \makecell[l]{Expand or collapse the process's\\ parameter fields} \\
\arrayrulecolor{gray!40}\hline
Click outside an open menu & \makecell[l]{Close the menu} \\
\arrayrulecolor{black}\hline\hline
\end{tabular}
\arrayrulecolor{black}
\caption{Keyboard and mouse controls in the \textsc{JUG} GUI.}
\label{tab: gui shortcuts}
\end{table}

\section{Fit Parity}
\label{app: parity}
  \begin{table*}[h]
      \centering
      \begin{tabular}{l c c c c}
          \hline\hline
          Parameter & \textsc{JUG} & \textsc{PINT} & $\sigma_\mathrm{JUG}$ & $|\Delta|/\sigma$ \\
          \hline
          \multicolumn{5}{c}{PSR~J1909$-$3744 \quad ($N_\mathrm{TOA}=35{,}037$, ELL1, GLS, RMS $=0.365\,\mu$s)} \\
          \hline
          ELONG    & $284.2208459$ & $284.2208459$ & $6.439\times10^{-10}$ & $0.0061$ \\
          ELAT     & $-15.15553321$ & $-15.15553321$ & $2.596\times10^{-9}$ & $0.0011$ \\
          PMELONG  & $-13.86991921$ & $-13.86991807$ & $5.483\times10^{-4}$ & $0.0021$ \\
          PMELAT   & $-34.31364843$ & $-34.31365324$ & $2.278\times10^{-3}$ & $0.0021$ \\
          PX       & $0.890994687$ & $0.8910300496$ & $7.751\times10^{-3}$ & $0.0046$ \\
          F0       & $339.3156923$ & $339.3156923$ & $9.622\times10^{-14}$ & $0.0000$ \\
          F1       & $-1.614808267\times10^{-15}$ & $-1.614808243\times10^{-15}$ & $5.615\times10^{-21}$ & $0.0044$ \\
          PB       & $1.533449452$ & $1.533449452$ & $7.796\times10^{-13}$ & $0.0023$ \\
          PBDOT    & $5.089342103\times10^{-13}$ & $5.089467189\times10^{-13}$ & $9.167\times10^{-16}$ & $0.0136$ \\
          A1       & $1.897991012$ & $1.897991012$ & $1.619\times10^{-8}$ & $0.0061$ \\
          XDOT     & $-1.95649777\times10^{-16}$ & $-1.958319478\times10^{-16}$ & $3.634\times10^{-17}$ & $0.0050$ \\
          M2       & $0.2087061874$ & $0.208699517$ & $9.326\times10^{-4}$ & $0.0072$ \\
          SINI     & $0.9980128482$ & $0.9980131322$ & $4.387\times10^{-5}$ & $0.0065$ \\
          TASC     & $56121.04496$ & $56121.04496$ & $8.496\times10^{-10}$ & $0.0086$ \\
          EPS1     & $4.448930721\times10^{-8}$ & $4.453678437\times10^{-8}$ & $7.901\times10^{-9}$ & $0.0060$ \\
          EPS2     & $-9.90179524\times10^{-8}$ & $-9.902859443\times10^{-8}$ & $4.624\times10^{-9}$ & $0.0023$ \\
          FD1      & $-8.745089653\times10^{-7}$ & $-8.744268619\times10^{-7}$ & $1.060\times10^{-8}$ & $0.0077$ \\
          JUMP1    & $-1.777923582\times10^{-5}$ & $-1.777941029\times10^{-5}$ & $1.761\times10^{-8}$ & $0.0099$ \\
          JUMP2    & $-1.949915552\times10^{-5}$ & $-1.949930813\times10^{-5}$ & $2.121\times10^{-8}$ & $0.0072$ \\
          \hline
          \multicolumn{5}{c}{PSR~J0437$-$4715 \quad ($N_\mathrm{TOA}=5{,}830$, DD, GLS, RMS $=0.398\,\mu$s)} \\
          \hline
          ELONG    & $50.46886858$ & $50.46886858$ & $2.533\times10^{-8}$ & $0.0005$ \\
          ELAT     & $-67.87322632$ & $-67.87322632$ & $1.989\times10^{-8}$ & $0.0012$ \\
          PMELONG  & $86.04772109$ & $86.04770265$ & $3.137\times10^{-2}$ & $0.0006$ \\
          PMELAT   & $-111.6714273$ & $-111.6713821$ & $3.396\times10^{-2}$ & $0.0013$ \\
          PX       & $9.700709869$ & $9.6992485$ & $9.468\times10^{-1}$ & $0.0015$ \\
          F0       & $173.687948$ & $173.687948$ & $2.816\times10^{-13}$ & $0.0000$ \\
          F1       & $-1.728365385\times10^{-15}$ & $-1.728365376\times10^{-15}$ & $2.211\times10^{-20}$ & $0.0004$ \\
          PB       & $5.741042383$ & $5.741042383$ & $1.551\times10^{-10}$ & $0.0005$ \\
          PBDOT    & $2.725374176\times10^{-12}$ & $2.725285613\times10^{-12}$ & $4.777\times10^{-13}$ & $0.0002$ \\
          A1       & $3.366740695$ & $3.366740695$ & $4.472\times10^{-8}$ & $0.0004$ \\
          XDOT     & $7.908466436\times10^{-14}$ & $7.908447701\times10^{-14}$ & $1.111\times10^{-15}$ & $0.0002$ \\
          ECC      & $1.929696849\times10^{-5}$ & $1.929693119\times10^{-5}$ & $3.901\times10^{-8}$ & $0.0010$ \\
          T0       & $58089.63373$ & $58089.63373$ & $1.514\times10^{-3}$ & $0.0015$ \\
          OM       & $2.324135007$ & $2.323991125$ & $9.495\times10^{-2}$ & $0.0015$ \\
          FD1      & $4.691555868\times10^{-5}$ & $4.691914799\times10^{-5}$ & $4.976\times10^{-7}$ & $0.0072$ \\
          FD2      & $-2.491638394\times10^{-5}$ & $-2.491872734\times10^{-5}$ & $5.566\times10^{-7}$ & $0.0042$ \\
          FD3      & $6.096304892\times10^{-6}$ & $6.096864915\times10^{-6}$ & $2.085\times10^{-7}$ & $0.0027$ \\
          JUMP1    & $4.372414228\times10^{-5}$ & $4.372405292\times10^{-5}$ & $1.668\times10^{-7}$ & $0.0005$ \\
          \hline\hline
      \end{tabular}
      \caption{Post-fit parameter comparison between \textsc{JUG} and \textsc{PINT} for PSR~J1909$-$3744 and PSR~J0437$-$4715, both fit to the NANOGrav 15-year dataset with a GLS fit including the noise model, using matched ephemeris, clock, and timescale conventions. Parameters are listed in the order given in the pulsar ephemerides. For each parameter we list the post-fit values, \textsc{JUG}'s $1\sigma$ uncertainty $\sigma_\mathrm{JUG}$, and the difference between them in units of that uncertainty, $|\Delta|/\sigma$. The per-epoch DMX dispersion parameters are held fixed and omitted. Every parameter agrees between the two packages to better than $0.014\sigma$.}
      \label{tab: app_jug_pint_fit_parity}
  \end{table*}

\bibliography{sample701}{}
\bibliographystyle{aasjournalv7}



\end{document}